\documentclass[]{aastex62}
\pdfoutput=1 

\usepackage{amsmath,amstext}
\usepackage[T1]{fontenc}
\usepackage{apjfonts} 
\usepackage[figure,figure*]{hypcap}
\usepackage{float}
\usepackage{rotating}
\usepackage{graphicx}
\usepackage{booktabs}
\usepackage{natbib}
\usepackage{multirow}

\definecolor{sof}{rgb}{0.4, 0.0, 0.6}

\newcommand{\SWIN}{Centre for Astrophysics \& Supercomputing, Swinburne University of Technology, Hawthorn, VIC 3122, Australia}

\shorttitle{SETI in the rETZ}
\shortauthors{Sheikh et al. 2020}

\begin{document}

\title{The Breakthrough Listen Search for Intelligent Life: A 3.95--8.00 GHz Search for Radio Technosignatures in the Restricted Earth Transit Zone}

\author[0000-0001-7057-4999]{Sofia Z. Sheikh}
\address{Department of Astronomy \& Astrophysics and Center for Exoplanets and Habitable Worlds\\ 525 Davey
Laboratory, The Pennsylvania State University, University Park, PA, 16802, USA}

\author[0000-0003-2828-7720]{Andrew Siemion}
\address{Department of Astronomy, University of California, Berkeley, 501 Campbell Hall 3411, Berkeley, CA, 94720, USA}
\affiliation{Department of Astrophysics/IMAPP, Radboud University, P.O. Box 9010, NL-6500 GL Nijmegen, The Netherlands}
\affiliation{SETI Institute, Carl Sagan Center, 189 Bernardo Ave., Mountain View CA 94043, USA}

\author[0000-0003-2516-3546]{J. Emilio Enriquez}
\address{Department of Astronomy, University of California, Berkeley, 501 Campbell Hall 3411, Berkeley, CA, 94720, USA}
\affiliation{Department of Astrophysics/IMAPP, Radboud University, P.O. Box 9010, NL-6500 GL Nijmegen, The Netherlands}

\author[0000-0003-2783-1608]{Danny C.\ Price}
\address{Department of Astronomy, University of California, Berkeley, 501 Campbell Hall 3411, Berkeley, CA, 94720, USA}
\affiliation{\SWIN}

\author[0000-0002-0531-1073]{Howard Isaacson}
\address{Department of Astronomy, University of California, Berkeley, 501 Campbell Hall 3411, Berkeley, CA, 94720, USA}

\author{Matt Lebofsky}
\address{Department of Astronomy, University of California, Berkeley, 501 Campbell Hall 3411, Berkeley, CA, 94720, USA}

\author[0000-0002-8604-106X]{Vishal Gajjar}
\address{Department of Astronomy, University of California, Berkeley, 501 Campbell Hall 3411, Berkeley, CA, 94720, USA}

\author[0000-0002-6221-5360]{Paul Kalas}
\address{Department of Astronomy, University of California, Berkeley, 501 Campbell Hall 3411, Berkeley, CA, 94720, USA}
\affiliation{SETI Institute, Carl Sagan Center, 189 Bernardo Ave., Mountain View CA 94043, USA}
\affiliation{Institute of Astrophysics, FORTH, GR-71110 Heraklion, Greece}

\begin{abstract}
We report on a search for artificial narrowband signals of 20 stars within the restricted Earth Transit Zone as a part of the ten-year Breakthrough Listen (BL) search for extraterrestrial intelligence. The restricted Earth Transit Zone is the region of the sky from which an observer would see the Earth transit the Sun with an impact parameter of less than 0.5. This region of the sky is geometrically unique, providing a potential way for an extraterrestrial intelligence to discover the Solar System. The targets were nearby (7--143 pc) and the search covered an electromagnetic frequency range of 3.95--8.00 GHz. We used the Robert C. Byrd Green Bank Telescope to perform these observations with the standard BL data recorder. We searched these data for artificial narrowband ($\sim$Hz) signals with Doppler drift rates of $\pm 20$ Hz s$^{-1}$. We found one set of potential candidate signals on the target HIP 109656 which was then found to be consistent with known properties of anthropogenic radio frequency interference. We find no evidence for radio technosignatures from extraterrestrial intelligence in our observations. The observing campaign achieved a minimum detectable flux which would have allowed detections of emissions that were $10^{-3}$ to $0.88$ times as powerful as the signaling capability of the Arecibo radar transmitter, for the nearest and furthest stars respectively. We conclude that at least $8\%$ of the systems in the restricted Earth Transit Zone within 150 pc do not possess the type of transmitters searched in this survey. To our knowledge, this is the first targeted search for extraterrestrial intelligence of the restricted Earth Transit Zone. All data used in this paper are publicly available via the Breakthrough Listen Public Data Archive (\url{http://seti.berkeley.edu/bldr2}).

\end{abstract}


\section{Introduction}
\label{sec: introduction}

The Search for Extraterrestrial Intelligence (SETI) is a sub-field of astrobiology that began in the early 1960s \citep{Cocconi1959, Dyson1960, Bracewell1960, Schwartz1961, Drake1961}, dedicated to the search for signatures of non-human technology or \textit{technosignatures} \citep{Tarter2006}. In the early days of the field, both the theory and scant observational campaigns primarily focused on searching for intentional communications in the radio wavelengths from extraterrestrial intelligences (ETIs). Radio wavelengths were favored based on low energy costs per bit, low signal attenuation in the galaxy from dust and gas, and theorized mutual predictability from prominent astrophysical emission lines \citep{Cocconi1959}. Since then, theoretical work has broadened the range of technosignatures that could be searched for (e.g. planetary atmospheric pollution \citep{Lin2014}, intentional laser signals \citep{Howard2004}, or transiting megastructures \citep{dyson1966search, Arnold2005}), but observational follow-up searches have been slow to follow.

Biosignatures and technosignatures are part of a continuum, recognizing the different ways that life can affect its environment and potentially be detectable within our own solar system or even at interstellar distances. The boom in biosignature research was spurred by the discovery of the first exoplanets \citep{wolszczan1992planetary, Mayor1995} and subsequent realization that potentially habitable environments are ubiquitous in the galaxy; 22\% of Sun-like stars harbor an Earth-size planet within their habitable zone \citep{petigura2013prevalence}, and there are \(0.16^{+0.17}_{-0.07}\) Earth-sized planets per M dwarf habitable zone \citep{dressing2015occurrence}.

Thousands of exoplanets have been discovered by Kepler \citep{Borucki2008} and TESS \citep{Ricker2014}, space-based facilities that use the transit method to find and confirm exoplanet candidates. This method relies on target systems with a favorable geometry; where the plane of the exoplanet's orbit around its host star is nearly aligned with the observing telescope. The periodic dimming of the lightcurves of the host stars provide information of the existence and basic properties of the exoplanets. With future instruments such as the James Webb Space Telescope \citep{Gardner2006}, we will be able to better determine the compositions of exoplanetary atmospheres, giving a way to remotely detect biosignatures and technosignatures on these transiting systems \citep{Lin2014, Krissansen-Totton2018}.

Systems in the Earth Transit Zone (ETZ) invert this geometry. As defined in \citet{heller2016search}, the ETZ is the projection of a band around Earth's ecliptic onto the celestial plane, where ETI observers could detect transits by Earth across the Sun. This band is between 0.520\(^\circ\) and 0.537\(^\circ\) wide, centered on the ecliptic, with the variance due to the eccentricity of Earth's orbit. The restricted Earth Transit Zone (rETZ) further limits this swath of the sky to a width of 0.262\(^\circ\) to only include stars that see Earth transit less than 0.5 solar radii from the Sun's center \citep{heller2016search}. 

The idea of the ecliptic as a preferred region for a SETI search has been around for a long time \citep{filippova1988ecliptic}. Some authors have compiled lists of stars with this preferential geometry \citep[i.e ][]{filippova1990list, filippova1992bioastronomy, castellano2004visibility}, while others have proposed searches of the region using both optical and radio telescopes \citep{shostak_villard_2004, conn2008seti}.

Other solar system planets have their own transit zones as well \citep{nussinov2009some}, the locations of which have been calculated in \citet{Wells2018}. Even if ETI observers were not serendipitously located in the Earth Transit Zone, they could reside in e.g. the Venus or Jupiter Transit Zones, and could thus potentially flag the solar system as an astrobiological object of interest. This widens the region of the sky that is useful to observe using the same logic as with the ETZ. 

ETIs that exist within the ETZ have many potential motivations to preferentially transmit a radio signal towards Earth, compared to ETIs that exist outside of the ETZ. For example, ETZ ETIs could detect Earth transit and know it has biosignatures in its atmosphere. Alternatively, they could observe humanity's transmissions of interplanetary radar, preferentially directed in our ecliptic. Even if they are not capable of detecting the Earth directly, ETZ ETIs could see the other planets in our solar system transit.  Finally, they could game-theoretically guess that
\textit{we} would guess to observe in the ETZ. Regardless of the motivation, radio transmission to a distant target is an expensive endeavour and is thus likely not to be omnidirectional. For the example reasons outlined above, and more, the ETZ is a region of the sky that may have a higher probability of targeted transmissions towards Earth. Therefore, it should be preferentially observed in SETI searches.

In this project, we designed our algorithms to search for a continuous fixed-amplitude, drifting, high signal-to-noise ratio (SNR) radio signal that has a narrowband ($\sim$Hz) frequency width. Astrophysically produced signals have a minimum width of hundreds of Hz due to natural and thermal broadening \citep{Cordes1997}. Human technology, meanwhile, has been able to easily create Hz-width signals for decades. Thus, narrowband searches can distinguish natural confounders in the radio sky from a SETI signal. In addition to narrowband signals, this work is also sensitive to some morphologies of broadband transmissions (as illustrated in Section \ref{ssec: rfi}), as well as strong unintentional signals (such as planetary radar).

\begin{figure}[ht]
    \centering
    \includegraphics[width=0.6\textwidth]{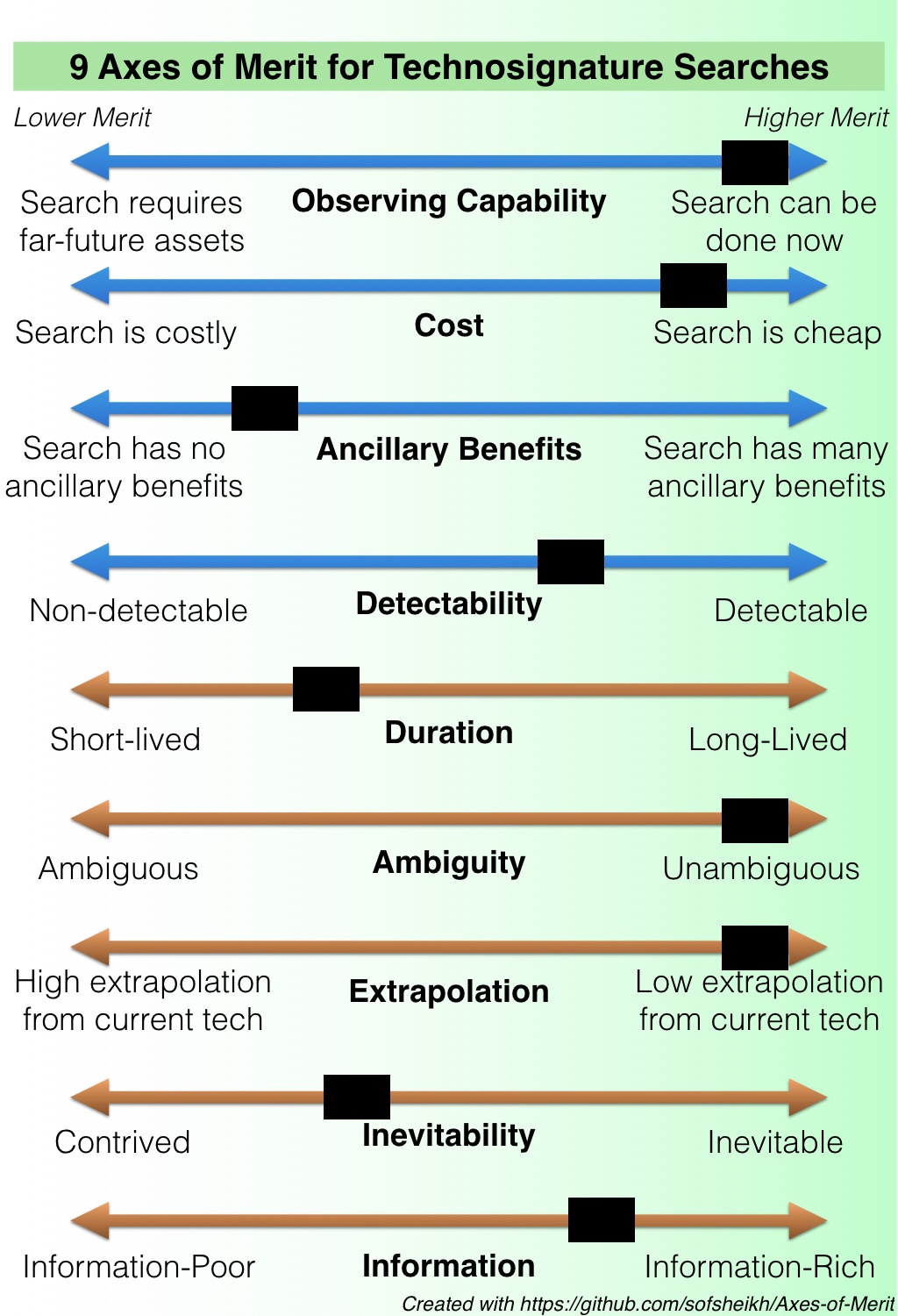}
    \caption{A qualitative ranking of the strengths and weaknesses of a narrowband search of the restricted Earth Transit Zone on the Nine Axes of Merit. The top four axes (blue) focus on practical/logistical concerns, while the bottom five (orange) focus on scientific concerns \citep{Sheikh2019b}}.
    \label{fig: aom}
\end{figure}

The motivation behind this search is well-characterized by the Nine Axes of Merit for Technosignature Searches \citep{Sheikh2019b}. Figure \ref{fig: aom} shows how the rETZ narrowband search ranks on each of the nine axes. This search scores well because: 

\begin{itemize}
    \item It can be done with existing observing capabilities (the Breakthrough Listen backend on the Green Bank Telescope, described in Section \ref{sec: observations}).
    \item It can be done cheaply with those capabilities.
    \item It is looking for a signal which would be relatively distinct against the astrophysical background.
    \item It is looking for a signal which has no known astrophysical confounders.
    \item It requires no extrapolation from existing technology on Earth.
    \item It could find a signal which carries some information from the ETI that transmitted it.
\end{itemize}

 The downsides to this search include the comparative lack of ancillary benefits, the reliance on the duration of the transmitter, and the assumption that an ETI will decide to send a narrowband signal to Earth. These particular observations are unlikely to produce data amenable to novel astrophysical interpretation due to the lack of radio emission at these locations and frequencies (which causes them to be desirable for SETI in the first place). However, radio SETI in general often develops and advances hardware that can be applied to other disciplines in observational astronomy \citep{macmahon2017recorder}. In addition, the strictness of the assumption of intentionality is ameliorated by the focus on the rETZ, where the existence of Earth would at least be known to the potential senders.

Given the background and motivation described above, we performed a search for artificial narrowband signals on 20 stars within the rETZ. Section \ref{sec: targetselection} explains how we selected and characterized these 20 targets. Section \ref{sec: searchparameters} defines the search parameters, delimiting the choices of sensitivity, drift rate, and signal strengths included in this work. Section \ref{sec: observations} describes the observing plan, instrumentation and observations themselves. Section \ref{sec: dataanalysis} presents data quality checks and the narrowband search analysis. Section \ref{sec: results} investigates the signals that were detected by our analysis. Section \ref{sec: discussion} and Section \ref{sec: conclusions} are the discussion and conclusion, respectively.

\section{Target Selection}
\label{sec: targetselection}

\begin{figure}[ht]
  \includegraphics[width=\textwidth]{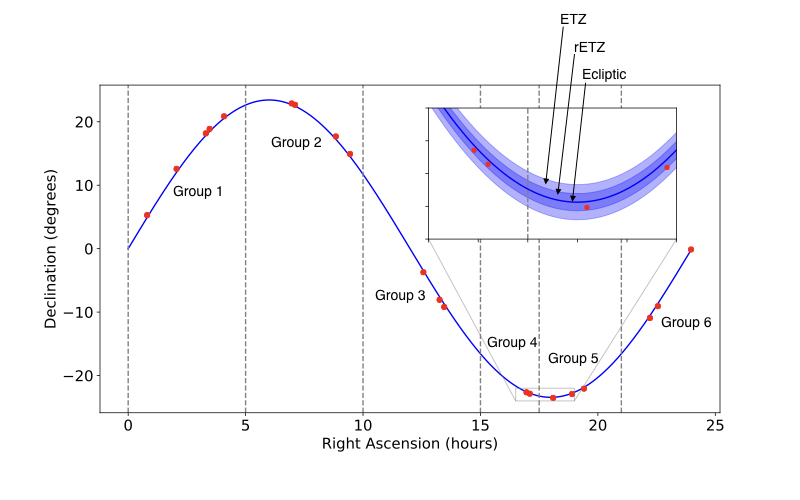}
  \centering
  \caption{The sample of 20 stars within the rETZ that were chosen for observation, plotted in red with their right ascension and declination. The inset figure shows the relationship between the ecliptic (solid blue line), the rETZ (darker shaded region in inset), and the ETZ (lighter shaded region in inset). All of the targets fall within the rETZ. We assigned observation group numbers by proximity in right ascension and used them to organize and streamline observations.}
  \label{fig: targets}
\end{figure}

We chose a sample of 20 stars from Table 1 of \citet{heller2016search}, containing 82 K and G dwarf stars within 1 kiloparsec of the Sun and inside the restricted Earth Transit Zone (rETZ). The chosen targets are shown in Figure \ref{fig: targets}, and detailed in Table \ref{tab: target_details}. We chose these targets from the original list solely to minimize their distance from Earth with three exceptions to homogenize our sample: our selection excludes HD 108754 and HD 123453, two known binaries \citep{pourbaix2004s, dommanget2002vizier}, and HD 121865, a red giant \citep{jones2011study}. The remaining targets are main-sequence stars that are billions of years old. We did not assign preference to known exoplanet hosts in our sample because the prevalence of exoplanets and the geometric constraints of exoplanet discovery methods imply that most stars without known planets are planet-hosting.

We divided the twenty targets into six groups based on their right ascension. A suitable flux calibrator, maser (for calibration in the frequency domain), and pulsar (for calibration in the time domain) were selected for each group (see Section \ref{ssec: dataquality}) and each group's observing script could be executed in a modular fashion.

Each target required a nearby off-source ``B-star'' that would be alternately observed with the main on-source ``A-star'' to identify radio frequency interference in the observations. While blank sky works as well for this purpose as an astrophysical target, using nearby stars as the off-source up to doubles the amount of SETI targets that can be observed in a single project. To minimize slewing time, the off-source positions were chosen to be within 7$^\circ$ of the primary on-source target. To ensure that potential signals from the on-source wouldn't also be caught in the off-source via sidelobes, we required the off-sources to be at least 5-6 beamwidths away from the on-source target. Given the distribution of targets along the ecliptic, 10 of the 20 targets were near enough to be on-source/off-source pairs, without the need for the selection of a new B-star. The other 10 off-source B-stars were chosen from the Breakthrough Listen target list based on the angular distance criteria previously described, but were not required to be nearby or within the rETZ. The distribution of targets within the groups, with their associated B-stars, is shown in Table \ref{tab: groups_and_times}.

Some of the chosen targets have been included in prior SETI searches. \citet{harp2016seti} observed HIP 15381, HIP 19054, HIP 33497, HIP 43418, and HIP 46339\footnote{\citet{harp2016seti} had a minimum detectable flux $\sim$50 times higher than this work, see Section \ref{sec: searchparameters}}. Breakthrough Listen has observed HIP 3764 \citep{enriquez2017breakthrough} and HIP 43418, HIP 111332, and HIP 46339 \citep{Price2019}. 

Most of the stars in the sample are solitary G or K dwarfs not known to host planets, but there are a few exceptions. HIP 83662 is a candidate proper motion binary\footnote{We count this as a single target because the binary is unconfirmed.} with a separation of 0".38 \citep{fabricius2002tycho}. HD 174995 is listed as a visual binary in the Tycho Double Star catalog\footnote{See previous footnote.} \citep{fabricius2002tycho}. Finally, HIP 109656 is the only exoplanet host in our sample, containing a 1.6 R$_{\oplus}$ planet with a 13-day orbital
period \citep{crossfield2016197}.

\begin{deluxetable}{lccccccccc}
\centering

\tablecaption{The list of 20 targets observed in this work. \label{tab: target_details}}
\tablehead{\colhead{ID} & 
           \colhead{RA (hr)} &
           \colhead{Dec (deg)} & 
           \colhead{Distance (pc)} & 
           \colhead{$\mu_{RA}$ (mas)} & 
           \colhead{$\mu_{Dec}$ (mas)} & 
           \colhead{V mag.} & 
           \colhead{Sp. Type} & 
           \colhead{\begin{tabular}[c]{@{}c@{}}EIRP$_{min}$\\(GW)\end{tabular}} &
           \colhead{\begin{tabular}[c]{@{}c@{}}EIRP$_{min}$\\($L_A$)\end{tabular}}}
\startdata
HIP 3765    & 00 48 23.0   & +05 16 50.2      & $7.4350^{+0.0049}_{-0.0049}$   & 755.6             & -1141.8           & 5.74$^{\beta}$       & K2.5V$^{\alpha}$      & 47    &  0.002\\
HIP 95417   & 19 24 34.2   & -22 03 43.8      & $27.6284^{+0.0476}_{-0.0476}$  & -230.8            & -451.6            & 10.899$^{\gamma}$    & K8V$^{\alpha}$        & 653   &  0.033\\
HIP 64688   & 13 15 30.8   & -08 03 18.5      & $40.7145^{+0.1079}_{-0.1079}$  & 49.5              & 58.7              & 8.06$^{\epsilon}$    & G5V$^{\delta}$       & 1417  &  0.071\\
HIP 34271   & 07 06 16.8   & +22 40 00.6      & $43.1941^{+0.0961}_{-0.0961}$  & -92.4             & -78.8             & 8.39$^{\xi}$          & G2V$^{\omicron}$      & 1595  &  0.080\\
HIP 33497   & 06 57 46.3   & +22 53 33.2      & $44.6959^{+0.1414}_{-0.1414}$  & -144.2            & -142.0            & 7.75$^{\xi}$          & G0$^{\zeta}$          & 1708  &  0.086\\
HIP 15381   & 03 18 20.0   & +18 10 17.8      & $47.4030^{+0.1074}_{-0.1074}$  & -83.1             & -103.8            & 7.540$^{\eta}$        & G0$^{\zeta}$          & 1921  &  0.096\\
HIP 9607    & 02 03 33.0   & +12 35 05.0      & $47.6268^{+0.0903}_{-0.0903}$  & 377.3             & -55.7             & 13.475$^{\epsilon}$  & K7V$^{\theta}$        & 1939  &  0.097\\
HIP 43418   & 08 50 36.9   & +17 41 21.5      & $50.1645^{+0.1004}_{-0.1004}$  & -158.4            & -61.2             & 9.51$^{\xi}$          & K0$^{\pi}$            & 2151  &  0.107\\
HIP 83662   & 17 05 59.6   & -22 51 24.3      & $50.3563^{+0.1270}_{-0.1270}$  & 39.8              & -325.7            & 10.00$^{\epsilon}$    & K2$^{\iota}$          & 2167  &  0.108\\
HD 174995   & 18 54 12.7   & -22 54 24.9      & $53.1762^{+0.1657}_{-0.1657}$  & -166.1            & -362.4            & 8.62$^{\gamma}$      & G9$^{\iota}$          & 2417  &  0.121\\
HIP 111332  & 22 33 21.5   & -09 03 48.8      & $61.0706^{+0.2268}_{-0.2268}$  & 297.7             & -61.0             & 8.86$^{\xi}$          & G3V$^{\delta}$        & 3188  &  0.159\\
HIP 118159  & 23 58 04.5   & -00 07 41.5      & $66.3060^{+0.2286}_{-0.2286}$  & -43.8             & -18.4             & 9.16$^{\xi}$          & G5Vn$^{\delta}$       & 3758  &  0.160\\
HIP 16136   & 03 27 55.3   & +18 52 56.4      & $66.8400^{+0.2216}_{-0.2216}$  & 14.3              & -60.6             & 8.48$^{\epsilon}$     & G0$^{\pi}$            & 3818  &  0.191\\
HIP 88631   & 18 05 46.7   & -23 31 03.8      & $84.1206^{+0.3474}_{-0.3474}$  & 25.9              & -212.7            & 9.28$^{\xi}$          & G6/8V$^{\kappa}$      & 6048  &  0.303\\
HIP 46339   & 09 26 49.4   & +14 55 40.7      & $85.6076^{+0.4690}_{-0.4690}$  & 21.9              & -92.6             & 8.38$^{\xi}$          & G0$^{\zeta}$          & 6264  &  0.313\\
HIP 109656  & 22 12 51.0   & -10 55 34.2      & 89.49\(^{+85}_{-44.951}\) \(^{\rho}\) & 180.4$^{\lambda}$ & -183.0$^{\lambda}$& 10.80$^{\epsilon}$    & K2V$^{\mu}$           & 6845  &  0.342\\
HIP 82986   & 16 57 30.2   & -22 38 37.1      & $127.4259^{+1.0830}_{-1.0830}$ & -3.1              & -30.1             & 9.78$^{\epsilon}$     & G0V$^{\kappa}$        & 13878 &  0.694\\
HIP 61349   & 12 34 13.3   & -03 43 16.2      & $136.6195^{+1.0807}_{-1.0807}$ & 13.4              & -57.8             & 8.57$^{\epsilon}$     & F5V$^{\delta}$        & 15952 &  0.798\\
HIP 65642   & 13 27 29.7   & -09 11 33.7      & $137.1272^{+1.4009}_{-1.4009}$ & -54.9             & -11.9             & 9.48$^{\epsilon}$     & G5V$^{\delta}$        & 16071 &  0.803\\
HIP 19054   & 04 04 56.5   & +20 51 23.3      & $143.4597^{+1.0373}_{-1.0373}$ & 12.8              & -52.1             & 8.98$^{\epsilon}$     & G0$^{\nu}$            & 17590 &  0.880\\
\enddata

\tablerefs{$^{\alpha}$\citet{Gray2006contributions}, $^{\beta}$\citet{vanbelle2009directly}, $^{\gamma}$\citet{zacharias2012usno}, $^{\delta}$\citet{houk1999michigan}, $^{\epsilon}$\citet{hog2000tycho}, $^{\zeta}$\citet{cannon1993vizier}, $^{\eta}$\citet{oja1987photometry}, $^{\theta}$ \citet{stephenson1986dwarf}, $^{\iota}$\citet{bidelman1985astronomical}, $^{\kappa}$\citet{houk1988michigan}, $^{\lambda}$\citet{prusti2016gaia}, $^{\mu}$\citet{dressing2017characterizing}, $^{\nu}$\citet{nesterov1995henry}, $^{\xi}$\citet{vanleeuwen2007hip}, $^{\omicron}$\citet{lockwood2009seasonal}, $^{\pi}$\citet{heckmann1975agk}, $^{\rho}$\citet{kunder2017rave}}
\tablecomments{For each target, we include its identifier (ID), right ascension in hours (RA), declination in degrees (Dec), distance in parsecs (Distance), proper motion in right ascension and declination in milliarcseconds ($\mu_{RA}$ and $\mu_{Dec}$), apparent visual magnitude, and spectral type. We also show minimum detectable Equivalent Isotropically Radiated Powers (EIRPs) for transmitters at each target (calculated in Section \ref{sec: searchparameters}). Column 9 gives this value in gigawatts and Column 10 gives this same value in units of $L_{A}$, where \(L_{A} = L_{Arecibo} = 20\)TW \citep{siemion20131}. All right ascensions, declinations, parallax distances, and proper motions are sourced from GAIA DR2 \citep{brown2018gaia} except where otherwise indicated.}
\end{deluxetable}

\begin{table}[ht]
    \centering
    \begin{tabular}{|c|c|c|c|}
    \hline
    Group & Target & Off-Source & Date Observed\\
    \hline
    \multirow{5}{*}{Group 1}
                    & HIP 3765  & HIP 4067 & 1 July 2017\\
                    & HIP 9607  & HIP 9848 & 1 July 2017\\
                    & HIP 15381 & $\downarrow$ & 1 July 2017\\
                    & HIP 16136 & $\uparrow$ & 1 July 2017\\
                    & HIP 19054 & HIP 19299 &1 July 2017\\
    \hline
    \multirow{4}{*}{Group 2}
                & HIP 33497  & $\downarrow$ & 3 November 2017\\
                & HIP 34271  & $\uparrow$ & 3 November 2017\\
                & HIP 43418 & HIP 43701 & 3 November 2017\\
                & HIP 46339 & HIP 46662 & 3 November 2017\\
    \hline
    \multirow{3}{*}{Group 3}
            & HIP 61349  & HIP 61711 & 30 June 2017\\
            & HIP 64688  & $\downarrow$ & 6 July 2017\\
            & HIP 65642 & $\uparrow$ & 6 July 2017\\
    \hline
    \multirow{3}{*}{Group 4}
        & HIP 82986  & $\downarrow$ & 6 July 2017\\
        & HIP 83662  & $\uparrow$ & 6 July 2017\\
        & HIP 88631 & HIP 88982 & 6 July 2017\\
    \hline
    \multirow{2}{*}{Group 5}
    & HIP 174995  & HIP 93072 & 6 July 2017\\
    & HIP 95417  & HIP 96440 & 6 July 2017\\
    \hline
    \multirow{2}{*}{Group 6}
    & HIP 109656  & $\downarrow$ & 1 July 2017\\
    & HIP 111332  & $\uparrow$ & 1 July 2017\\
    & HIP 118159  & HIP 1 & 1 July 2017\\
    \hline
    \end{tabular}
    \caption{Targets broken into observing groups with their chosen off-source B-star. Pairs of targets with up and down arrows in the table were used as each other's off-sources. Dates for each observation are also shown.}
    \label{tab: groups_and_times}
\end{table}

\section{Search Parameters}
\label{sec: searchparameters}

Observations were performed using the GBT C-band receiver (3.95 -- 8.00 GHz). We chose to observe in C-band because it is comparatively underexplored in SETI programs, has less interference than other regions of frequency, is already integrated with the Breakthrough Listen hardware, and is a wide bandwidth receiver on the GBT (strengthening the Figure of Merit of the search, see Section \ref{sec: discussion}). The System Effective Flux Density (SEFD) can be calculated from the gain and system temperature of this receiver given in the GBT Proposer's guide (\citep{GBTSuppor2017} as shown in Equation \ref{eq: sefd}.

\begin{equation}
    SEFD = \frac{T_{\mathrm{sys}}}{G}
    \label{eq: sefd}
\end{equation}

Here, \(\rm{T_{sys}}\) is system temperature in Kelvin, \(G\) is the gain of the receiver, and the SEFD is measured in Jy. The radiometry equation can then be used to derive an equation for the minimum detectable flux (see \citealt{siemion20131}), shown in Equation \ref{eq: min_flux}.

\begin{equation}
    F_{\rm{min}} = \sigma_{\rm{threshold}} \times SEFD \times \sqrt{\frac{\Delta \nu }{T \sqrt{2}}}
    \label{eq: min_flux}
\end{equation}

In Equation \ref{eq: min_flux}, \(\sigma_{threshold}\) defines the signal-to-noise ratio required for a signal to be treated as a potential candidate, \(\Delta \nu\) is the frequency resolution of the observation, and \(T\) is the on-source integration time. The factor of $\sqrt{2}$ comes from an observation with two polarizations.

For this study, we chose a threshold signal-to-noise value of 10, an on-source integration time of 288 seconds, and a 2.7 Hz frequency resolution\footnote{This is the standard output resolution for Breakthrough Listen's High Spectral Resolution data product \citep{Lebofsky2019}}. Thus, the minimum detectable flux is \(7.14 \times 10^{-26}\) Wm$^{-2}$. We can then find the minimum Equivalent Isotropically Radiated Power (EIRP) to quantify the strength of the transmitter that this search is sensitive to, by multiplying the minimum detectable flux above by \(4 \pi d^2\), where $d$ is the distance to the source. We find that a minimum transmitter EIRP of 47\,GW would be required for a detection from the nearest star in the target list (HIP 3765) and a minimum of 18 TW would be required for a detection from the furthest star in the target list (HIP 65642). EIRPs for all targets are shown in Table \ref{tab: target_details}.

We used the Python-based software \textit{turboSETI}, developed by Breakthrough Listen \citep{enriquez2017breakthrough}, to search the filterbank data for narrowband SETI signals. \textit{TurboSETI} requires the user to input two parameters before searching the data: \(SNR_{threshold}\) (chosen to be 10 as described above) and \(\frac{\Delta \nu}{\Delta t}_{max}\), the maximum drift rate, in Hzs$^-1$, that the algorithm will search. A Doppler drift rate is a spectral shift over the course of an observation that occurs because of relative radial acceleration between the telescope and the transmitter. We chose a maximum drift rate of 20 Hz/s, which, when divided by the maximum and minimum frequencies in the search, corresponds to a normalized\footnote{For a more thorough discussion of the normalized drift rate, see 
\citep{Sheikh2019a}.} range of drift rates from 2.5 nHz to 5 nHz\footnote{1 nHz = a drift of 1 Hz/s at 1 GHz}. The normalized drift rate from an transmitter on the surface of Io is 2.39 nHz \citep{Sheikh2019a}; the initial choice of 20 Hz/s would be sufficient to detect, at the maximum-possible SNR, a transmitter on the surface of any planet, moon, or asteroid in the solar system.

\section{Observations}
\label{sec: observations}

Observations for this study were performed as part of the \textit{Breakthrough Listen Initiative}, the most comprehensive data-driven search for technosignatures ever performed \citep{Worden2017}. The project saw first light on its radio observations in 2015 and data analysis has been ongoing \citep{enriquez2017breakthrough, isaacson2017breakthrough, Price2019}. 

This rETZ observing campaign was allotted twelve hours of Breakthrough Listen time on the GBT, with a \(20\%\) overhead time set aside for calibration and slewing of the telescope. We implemented a flux-limited method such that every target received an equal amount of observing time (28.8 minutes). The differences in distance in the sample made a luminosity-limited method impractical; a luminosity-limited survey would require \(\sim 400 \times\) more observing time for the furthest star than the nearest star. 

Each target was observed in an ABABAB pattern: a single target was observed for 288 seconds, then a nearby off-source location was observed for another 288 seconds. This process was repeated a total of three times. The choice of the off-source is discussed in Section \ref{sec: targetselection}. On-off observing excises ground-based RFI by removing signals that are present in both the A-star and the B-star. A signal only present in the A-star or the B-star, on the other hand, indicates a signal that is fixed on the sky.

We used the Breakthrough Listen digital data recorder for these observations. In this configuration, data are initially acquired as voltages written to disk using the VEGAS instrument \citep{Prestage2015} on the GBT.  Voltage data are typically discarded, but can optionally be retained. The back-end divides the digitized signals into 512 frequency channels with an 8X overlapped polyphase filter bank, records at 8 bits per sample, and is currently capable of recording up to 12 GHz of bandwidth. Subsequently, the data are processed to produce Stokes I (total power) spectral data products at multiple resolutions. For more information, see \cite{macmahon2017recorder}. 

Each group of targets (shown in Figure \ref{fig: targets}) had its own \textsc{Astrid} observing script. We performed the observations on four different nights throughout 2017. Observation dates are shown in Table \ref{tab: groups_and_times}. Unfortunately, a total of 8 of the 20 targets (consisting of the first half of Group 3, in addition to Group 1 and Group 5) had shifted tunings, such that the fourth and final bank was skewed towards frequencies higher than the receiver could record. For these targets, the recorded frequency range was $\sim$ 5--8 GHz instead of 4--8 GHz.

\section{Data Analysis}
\label{sec: dataanalysis}

\subsection{Data Quality}
\label{ssec: dataquality}

\begin{table}[ht]
    \centering
    \begin{tabular}{|c|c|c|c|c|c|c|}
    \hline
                    & Group 1 & Group 2 & Group 3 & Group 4 & Group 5 & Group 6 \\
                    \hline
    Flux Calibrator & 3C 48 & 3C 123 & 3C 286 & 3C 353 & DR 21 & NGC 7027\\
    Pulsar          & B0031-07 & B0950+08 & B1133+16 & B1642-03 & B1929+10 & J2145-0750 \\
    Maser           & S235 & S231 & W3OH & 35.03+0.35 & 42.03+0.19 & NGC 7538 \\
    \hline
    \end{tabular}
    \caption{The calibrators that were observed to check data quality. Every observing group had its own pulsar, maser, and flux calibrator in order to assure that all systems were functioning correctly during the subsequent SETI observations.}
    \label{tab: calibrators}
\end{table}

Before every target group, three calibration sources were observed: a flux calibrator, a maser, and a pulsar. These calibrators are listed in Table \ref{tab: calibrators}. The flux calibrators were used to calibrate the maser flux densities. We chose these sources based on their proximity to the central right ascension of their respective target group and their high spectral flux densities. We observed each calibrator with the GBT C-band receiver for 120 seconds. 

We used the observed astrophysical calibrators (masers and pulsars) as operability tests; we are not using them as flux calibrators or otherwise to assess sensitivity, but rather taking the sensitivity to be flat across the band at the value given in the GBT Observer's Guide \citep{GBTSuppor2017}. We applied the \textsc{PRESTO} utility \textsc{prepfold} for pulsar folding and candidate optimization \citep{Ransom2011} to the High Time Resolution data product (\(\Delta t\) = 349 \(\mu\)s) \citep{Lebofsky2019} from each pulsar observation. Five of the six pulsars were detected in our observations, with the exception of J2145-0750 which appears to have strong frequency and time dependent RFI across the bandpass. We also produced time-averaged spectra for the maser observations and searched them for maser emission in methanol (CH\(_3\)OH) at 6668.5192 MHz \citep{Breckenridge2002}. Five of the six masers were detected in our observations, with the exception of S235 in Group 1. We determined that we had mistakenly used the optical coordinates for this maser, and the offset of $\sim 0.2$ beamwidths could have contributed to the non-detection. The Group 1 pulsar, B0031-07, was detected without issue. We thus determined that the system was working as expected.

\subsection{Candidate Selection}
\label{ssec: candidates}

We searched the filterbank data for SETI signals using \textit{turboSETI}. All filterbank files used in this analysis are publicly available via Breakthrough Listen Data Release 2 (\citet{lebofsky2020datarelease}, in prep). The data were searched for hits with an incoherent tree dedispersion algorithm (first developed by \citet{taylor1974sensitive}) implemented in \textit{turboSETI} which searches for narrowband Doppler-drifted signals. For more details on this algorithm, see \citet{enriquez2017breakthrough}. The 90 individual 5-minute observations were searched for hits above a signal-to-noise ratio (SNR) of 10, up to a maximum absolute drift rate value of 20 Hzs$^{-1}$. With a minimum Doppler drift rate of about 0.01 Hzs$^{-1}$, this results in approximately 4000 trial drift-rates for each 2.7 Hz frequency channel in each observation. Once completed, this process created a list of 425168 detections above the SNR threshold, or ``hits''. These hits were stored in plaintext files containing information such as the drift rate of the hit, the SNR, and the central frequency. The plaintext files were read into Python using the \textsc{PANDAS} package for the next stage of analysis with \textit{turboSETI} and processed into plots with the filterbank format reader from the Breakthrough Listen I/O Methods for Python (\textsc{blimpy}) package \citep{Price2019blimpy}.

To determine which hits were associated with one another, and which were potentially persistent and localized on the sky, we grouped them into 1-3 hit ``events''. These events were created by choosing an initial hit and then searching for any hits in later ON observations between the central frequency of the first observation and the frequency obtained by tracking twice the measured drift rate over the elapsed time. For each event, the OFF observations are subsequently searched for the same signal -- if it is found, the signal is not localized on the sky and is discarded. We performed this filtered search using the \textsc{find\_event} method in \textit{turboSETI}. 

Filtered event searches must strike a balance between inclusivity -- a SETI-search algorithm should not throw out potential candidates -- and selectivity -- the size of the pool of events should be contingent on the searchers' resources to visually inspect candidates. To strike this balance, we applied two different filters to select events from our total pool of hits: All-Hit and Only Non-Zero Drift.

For an event to pass the All-Hit filter, it must have a hit in all three ON observations, with the placement of subsequent ON hits dictated by the propagation of the measured drift rate in the first ON observation. All hits must have an SNR of at least 10, though they may have any drift rate including zero.

For an event to pass the Only Non-Zero Drift filter, it must have a hit in just one of the three ON observations. All hits must have an SNR of at least 10, and, for this filter-level, they must have a non-zero drift rate in the topocentric frame. We chose the non-zero drift rate cut because leaving zero drift signals at this filter-level creates too many candidates to manually vet, and signals with zero drift rate are characteristic of ground-based radio frequency interference (RFI).

For both filters, all events were saved to output dataframes and dynamic spectra were produced from each event. Stars that were the A-star in the ABABAB observing pattern were compared to all three subsequent B-star observations. For technical reasons, the data from the B-stars were most easily analyzed using a BABAB scheme. Having only 2 OFF positions slightly decreases our discrimination to RFI, but does not affect our sensitivity and has no influence on our top candidate list. This leads to some composite dynamic spectra in Figures \ref{fig:rfi}--\ref{fig: cand_four} having six panels, and others having five. Some RFI-heavy observations for the Only Non-Zero Drift filter, by the SNR cut of 10, produced thousands of events. This would have resulted in thousands of plots for a single target; in these cases, we raised the SNR threshold for plotting (up to 20--100) to reduce the number of plots to about 100, which could all be visually inspected. In these cases, we confirmed via the hit table that all of the events, even those that were not plotted, had similar drift rates, frequencies, and SNRs. We acknowledge that a real SETI signal with the same hit characteristics as the local RFI could be thrown out by this crude methodology, but consider the probability of such a situation low enough to ignore for this study.

\begin{figure}[ht]
    \centering
    \includegraphics[width=0.8\textwidth]{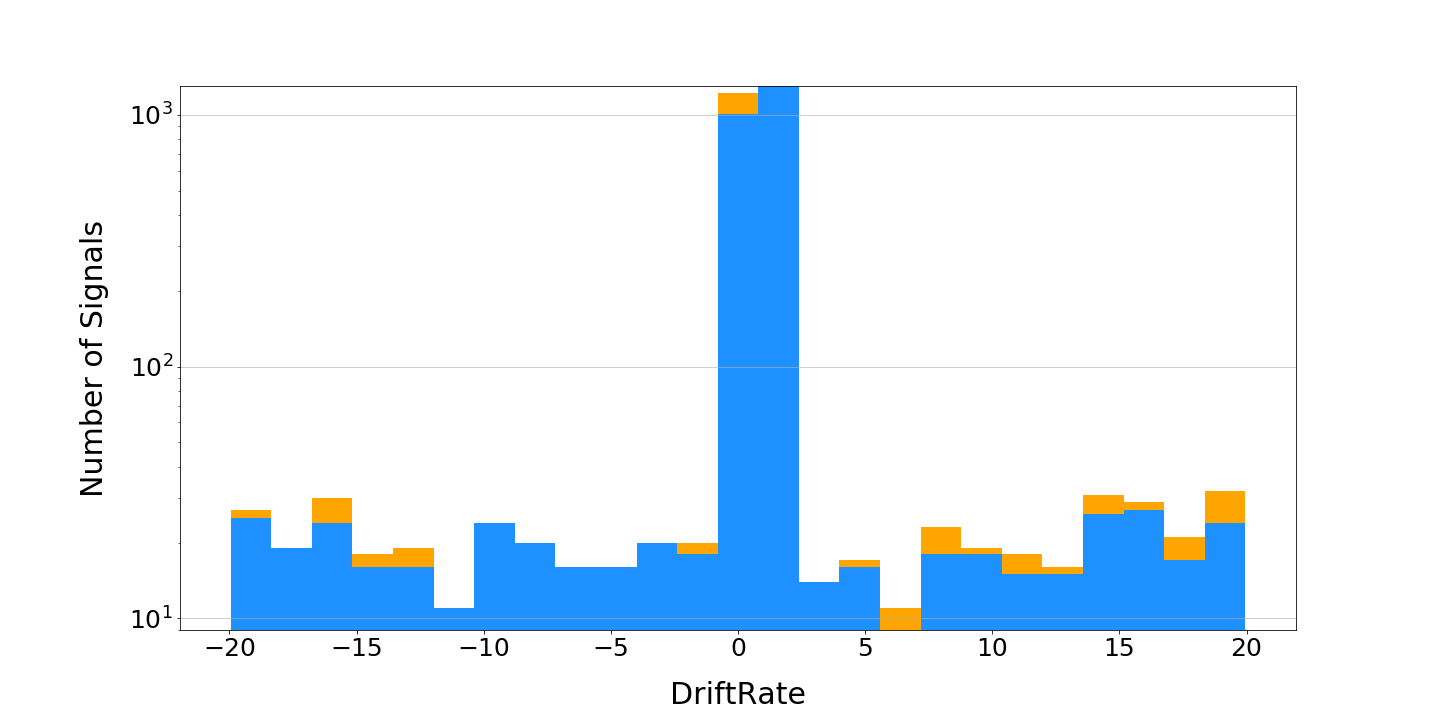}
    \label{fig: drhist}
    \caption{Logarithmic stacked-bar histogram of the drift rates of the events found by the ETZ survey. The Only Non-Zero Drift filter is shown in blue (bottom bars) and the All-Hit filter is shown in orange (top bars). The chosen drift rate range for the search was $\pm 20$ Hzs$^{-1}$, and events for each filter were defined as described in Section \ref{ssec: candidates}. Events tend to be clustered around zero drift rate, even though hits with exactly zero drift rate were removed with the Only Non-Zero Drift filter. This behaviour is expected in a set of events dominated by radio frequency interference because radio frequency interference produced in the same frame as the telescope will have a low or zero drift rate.}
\end{figure}

\begin{figure}[ht]
    \centering
    \includegraphics[width=0.8\textwidth]{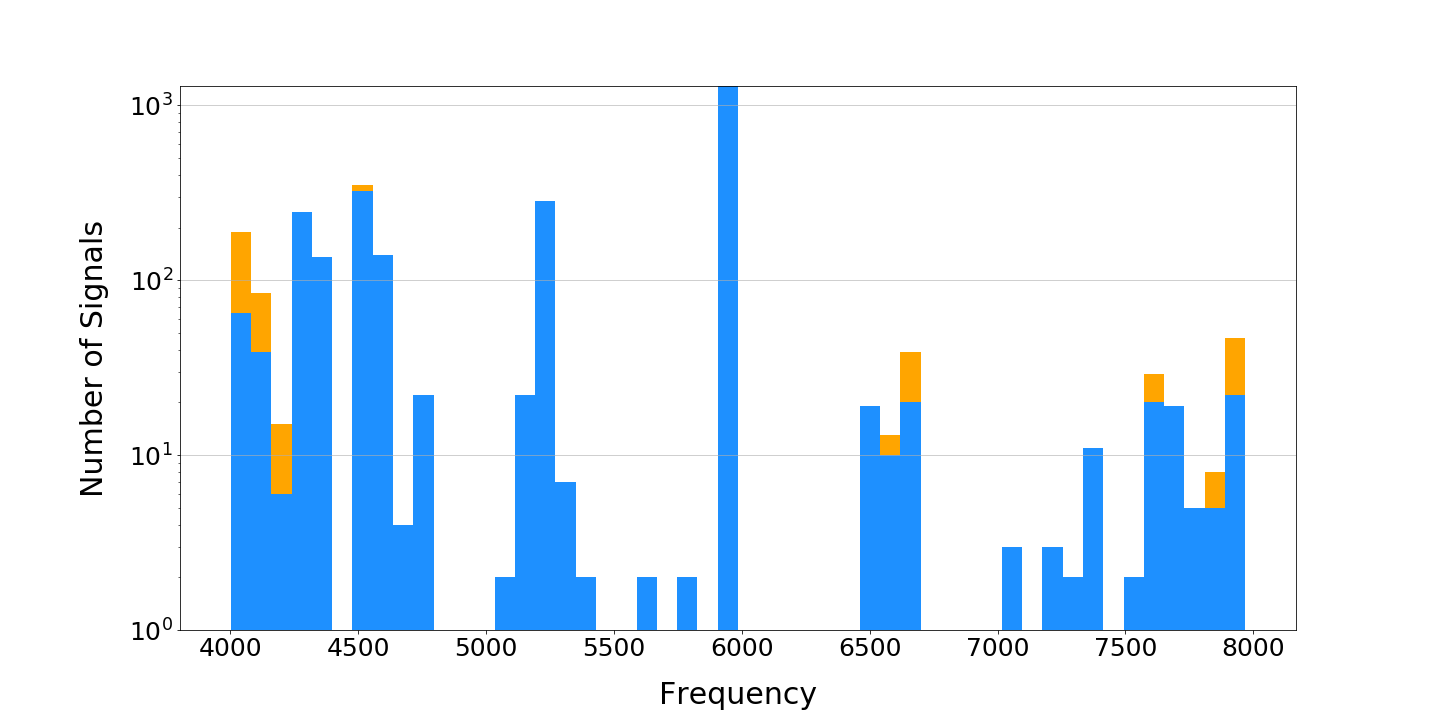}
    \caption{Logarithmic stacked-bar histogram of the frequencies of the events found by the ETZ survey. The Only Non-Zero Drift filter is shown in blue (bottom bars) and the All-Hit filter is shown in orange (top bars). The displayed frequency interval is 4--8 GHz. Events were defined as described in Section \ref{ssec: candidates}. Events appear throughout the band, with a large excess just below 6 GHz. This aligns with hundreds of FCC allocated telecommunications bands whose frequency ranges start at 5.925 GHz.}
    \label{fig: freqhist}
\end{figure}

\begin{figure}[ht]
    \centering
    \includegraphics[width=0.8\textwidth]{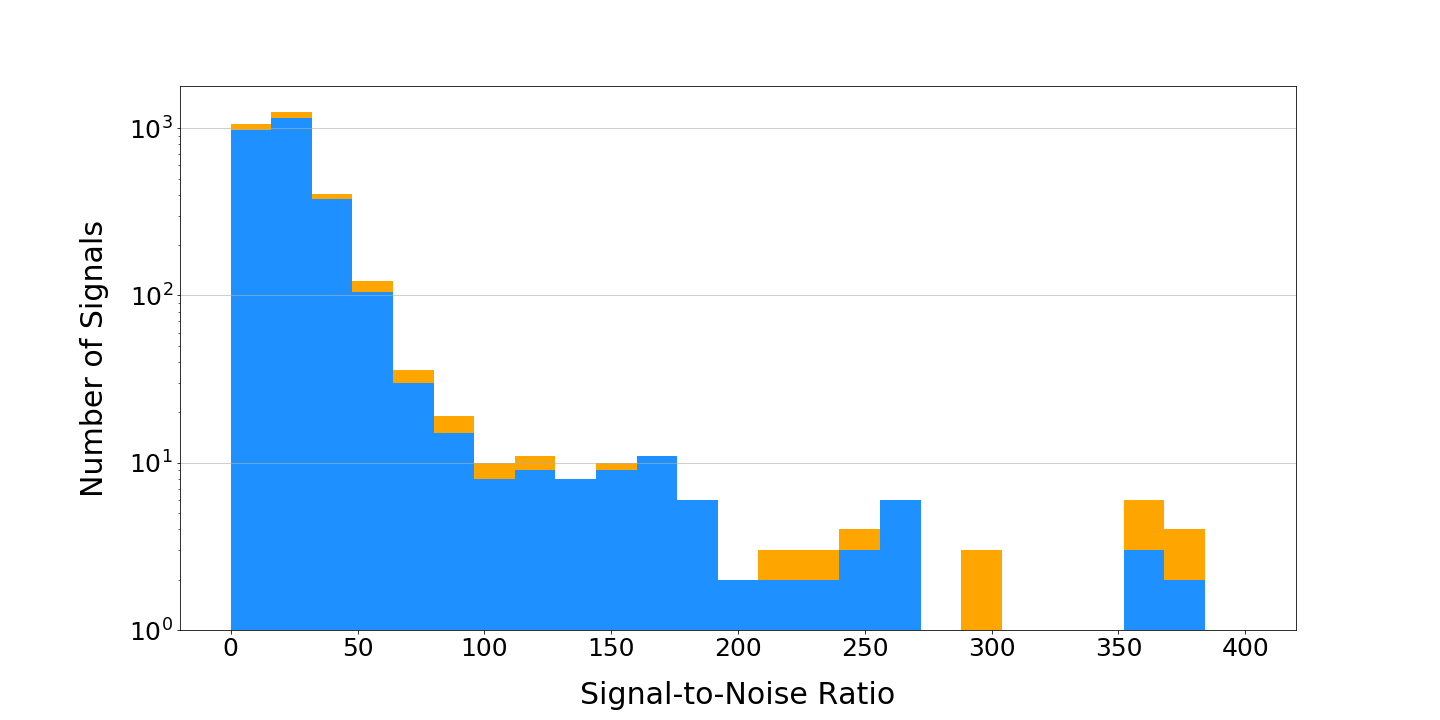}
    \caption{Logarithmic stacked-bar histogram of the signal-to-noise-ratios of the events found by the ETZ survey. The Only Non-Zero Drift filter is shown in blue (bottom bars) and the All-Hit filter is shown in orange (top bars). The chosen SNR cut for the search was 10, and events were defined as described in Section \ref{ssec: candidates}. Most events are weak, appearing very close to the threshold on the left side of the plot, and strong events are less common.}
    \label{fig: snrhist}
\end{figure}

We define an ``event-related hit'' as a hit which is part of an event. There can be up to three event-related hits in the same event -- three hits related by drift rate that appear in all three ON observations. On the Only Non-Zero Drift filter, however, hits are flagged even if they do \textit{not} appear in all three ON observations, so some events may be comprised of only one or two hits. The All-Hit filter produced 260 event-related hits, while the Non-Zero Drift filter produced 2726 event-related hits. Histograms showing the characteristics of the population from both filters are shown in Figures \ref{fig: drhist}, \ref{fig: freqhist}, and \ref{fig: snrhist}.

\section{Results}
\label{sec: results}

We visually inspected the 2986 event-related hits that were found with the All-Hit and Only Non-Zero Drift filters (Section \ref{ssec: candidates}). The majority were clearly radio frequency interference (RFI) due to the presence of a continuation of the same signal-to-noise ratio (SNR) signal in the on-source and the off-source. Most event plots fell into various morphology classes of RFI, discussed in Section \ref{ssec: rfi}. Only 103 event plots, grouped into four sets, warranted further inspection after this visual cut. These events are discussed in Section \ref{ssec: bestcand}. 

\subsection{Radio Frequency Interference}
\label{ssec: rfi}

\begin{figure}[ht]
\gridline{\fig{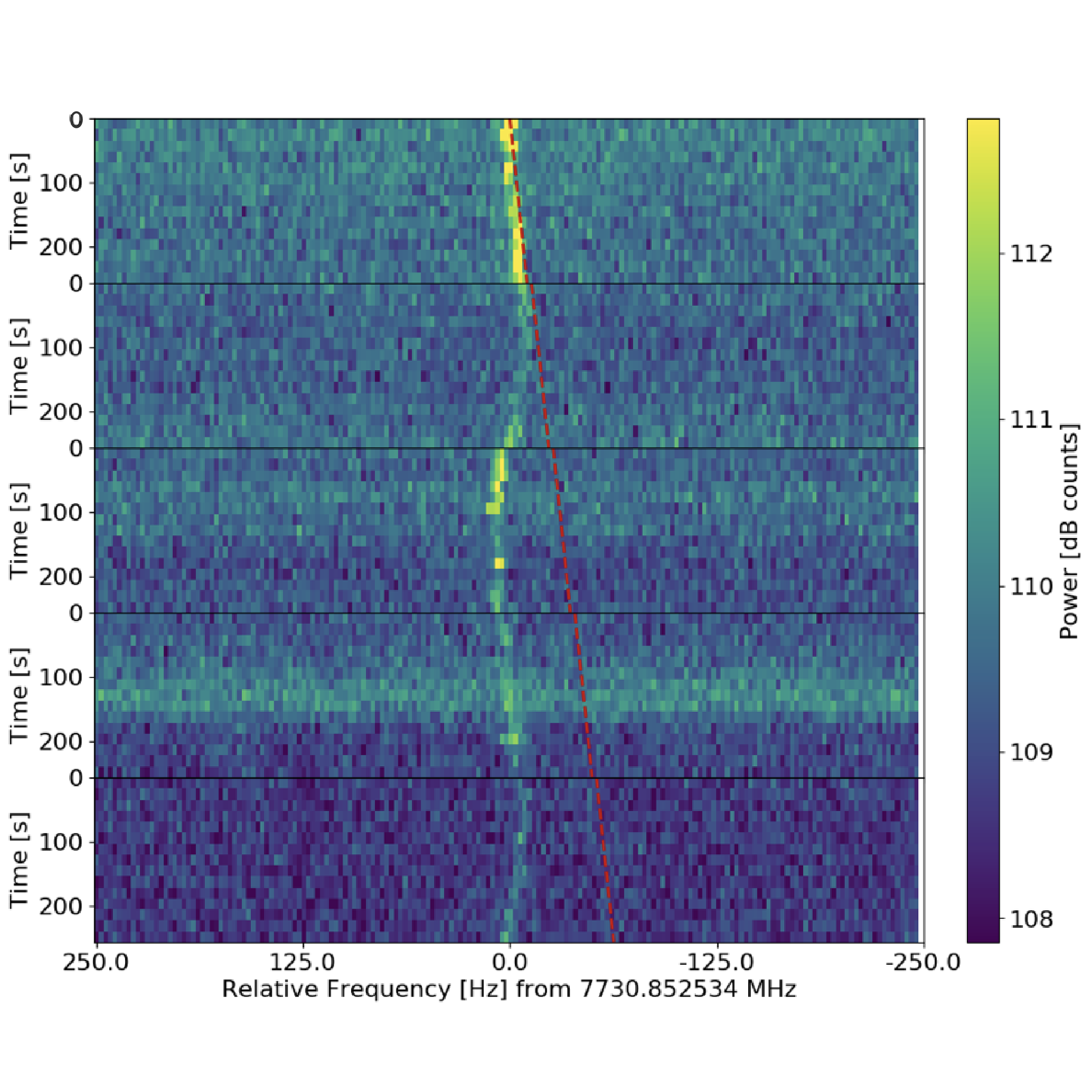}{0.3\textwidth}{(a)\label{fig:snake}}
          \fig{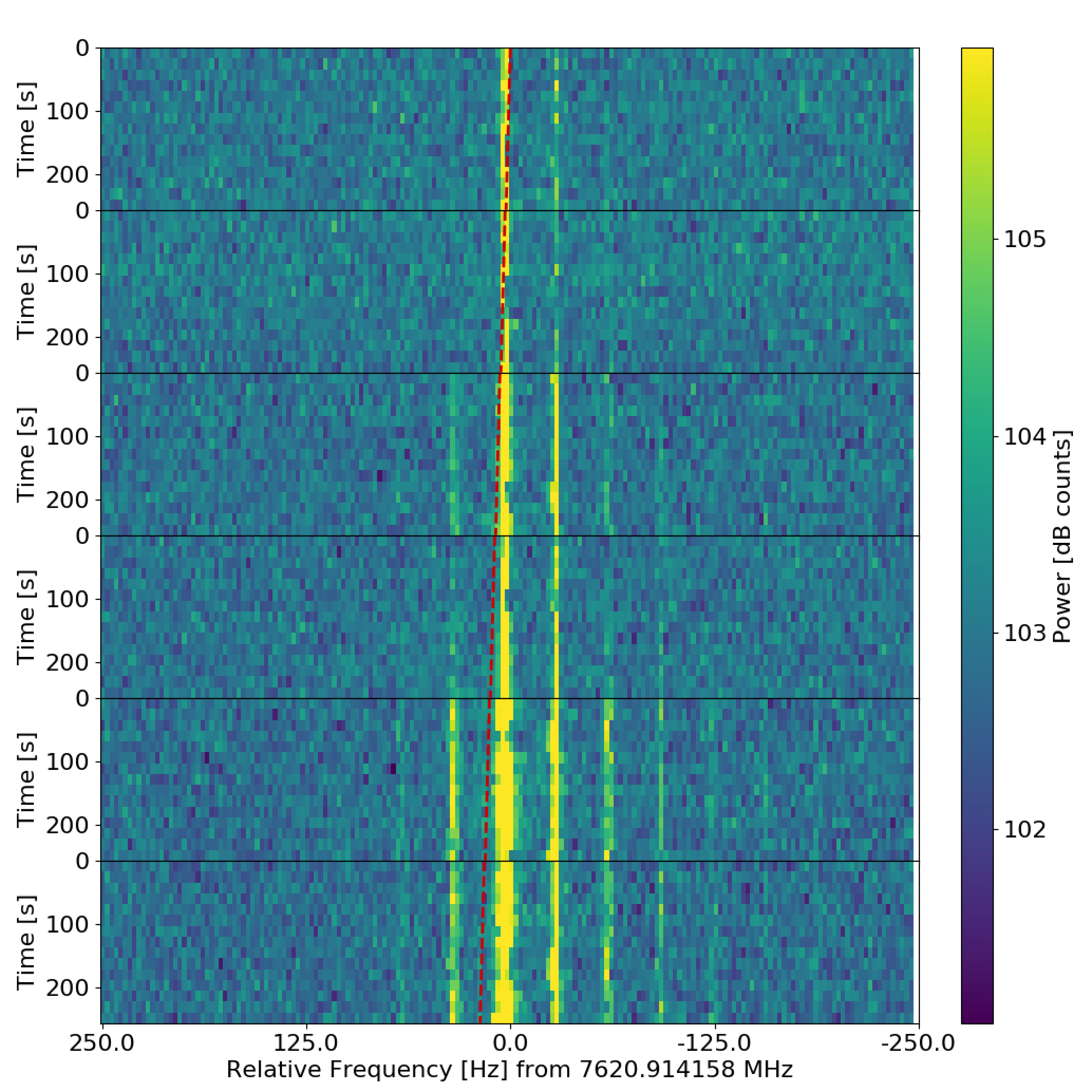}{0.3\textwidth}{(b)\label{fig:comb}}
          \fig{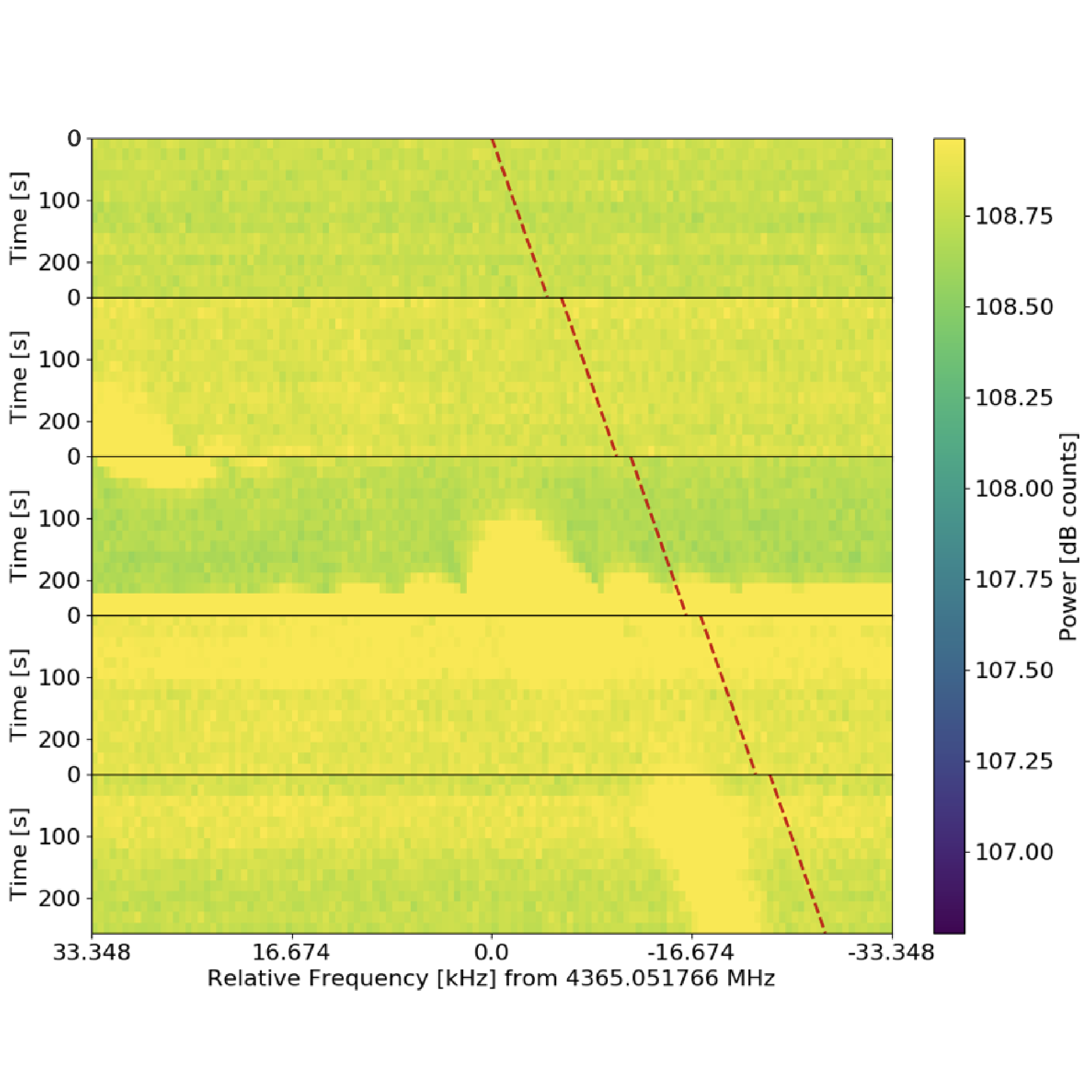}{0.3\textwidth}{(c)\label{fig:wash}}}
\gridline{\fig{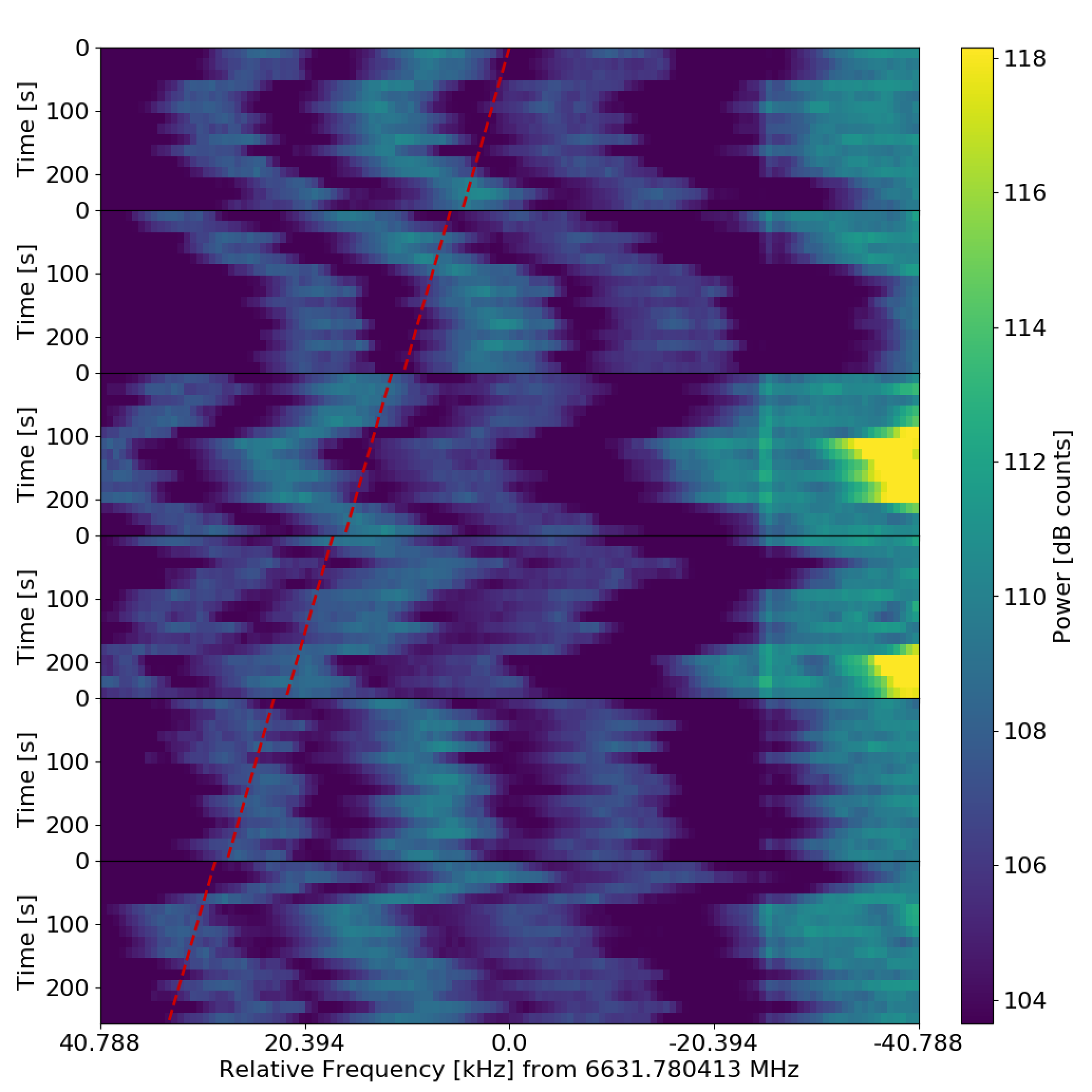}{0.3\textwidth}{(d)\label{fig:wave}}
          \fig{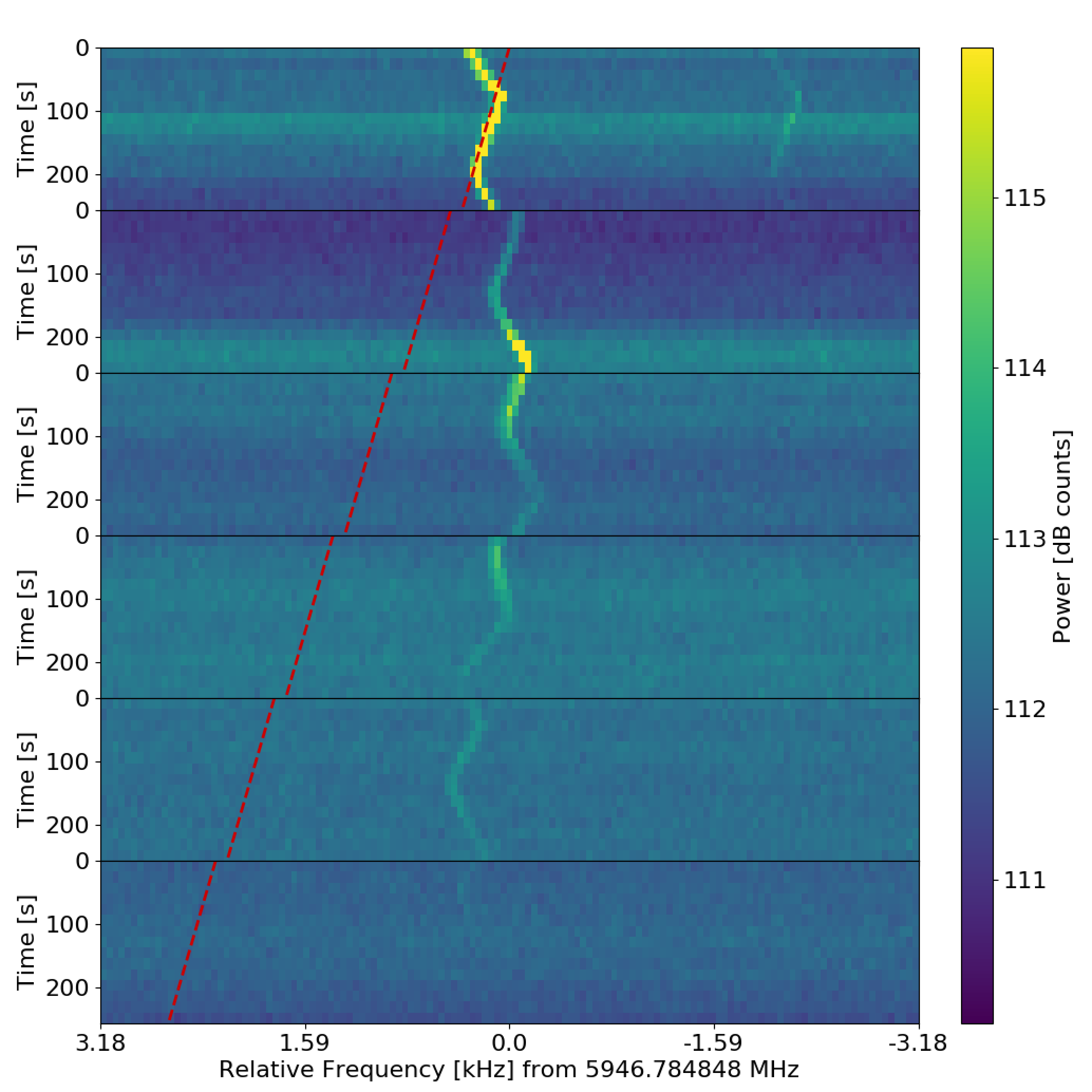}{0.3\textwidth}{(e)\label{fig:squiggle}}
          \fig{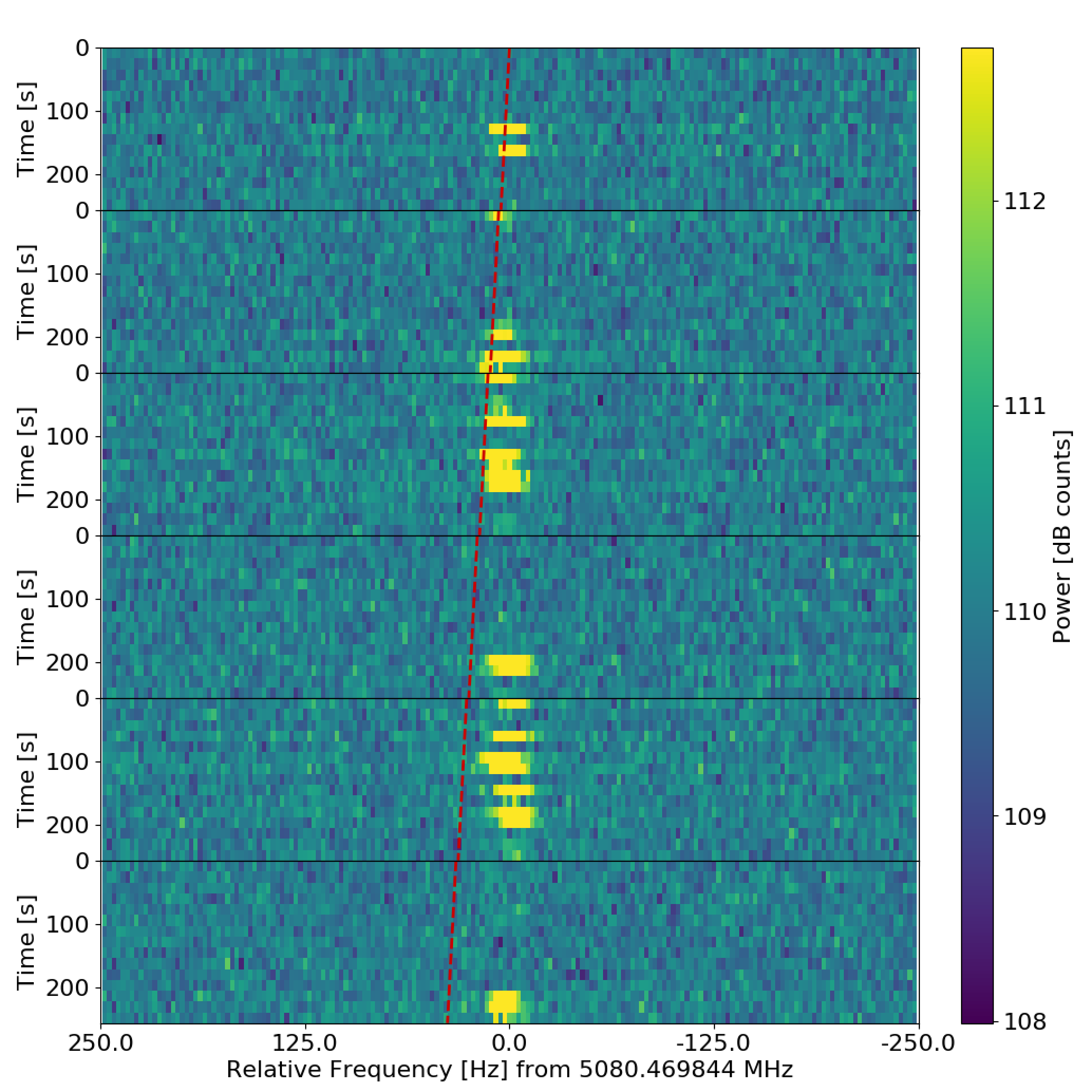}{0.3\textwidth}{(f)\label{fig:pulse}}}
\gridline{\fig{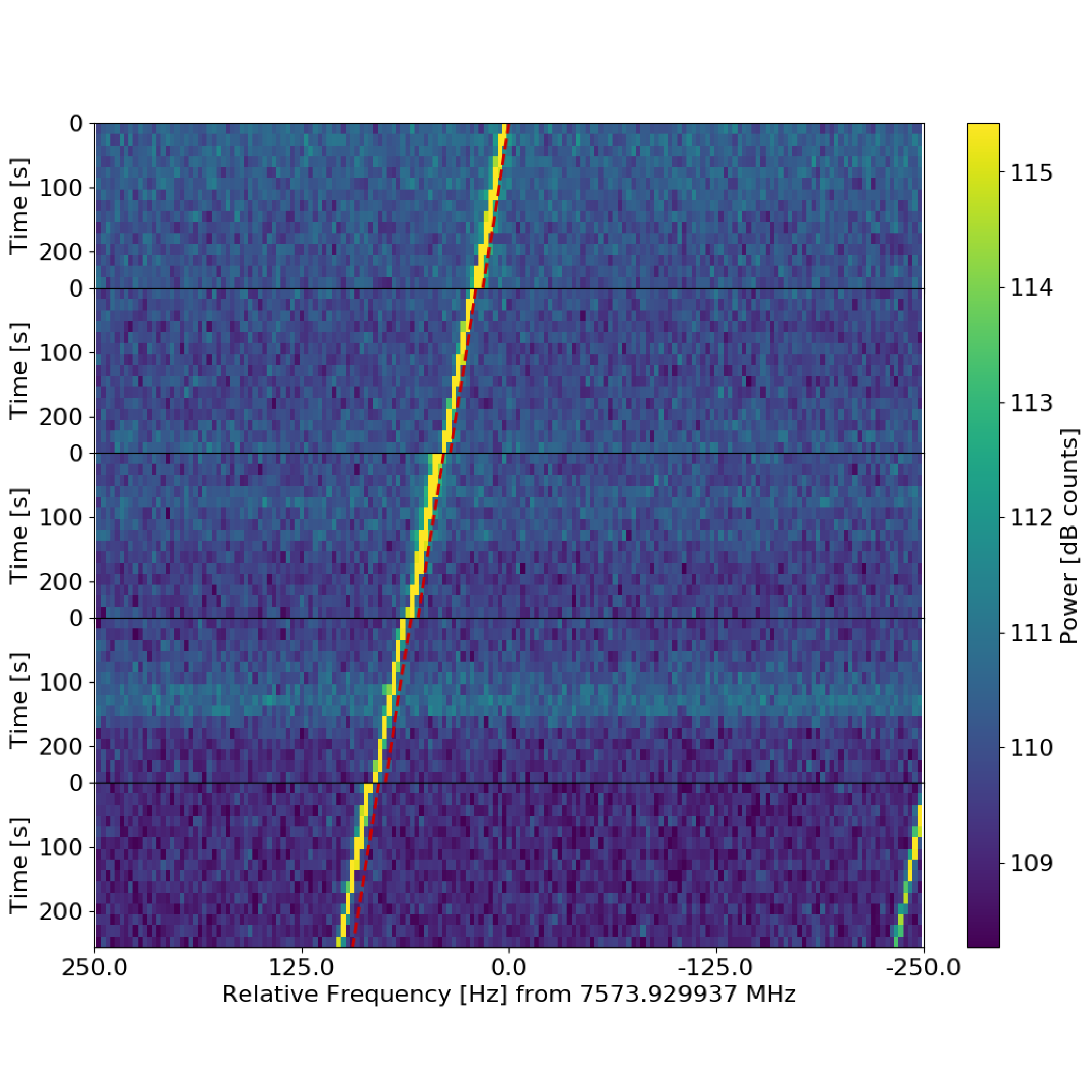}{0.3\textwidth}{(g)\label{fig:line}}
          \fig{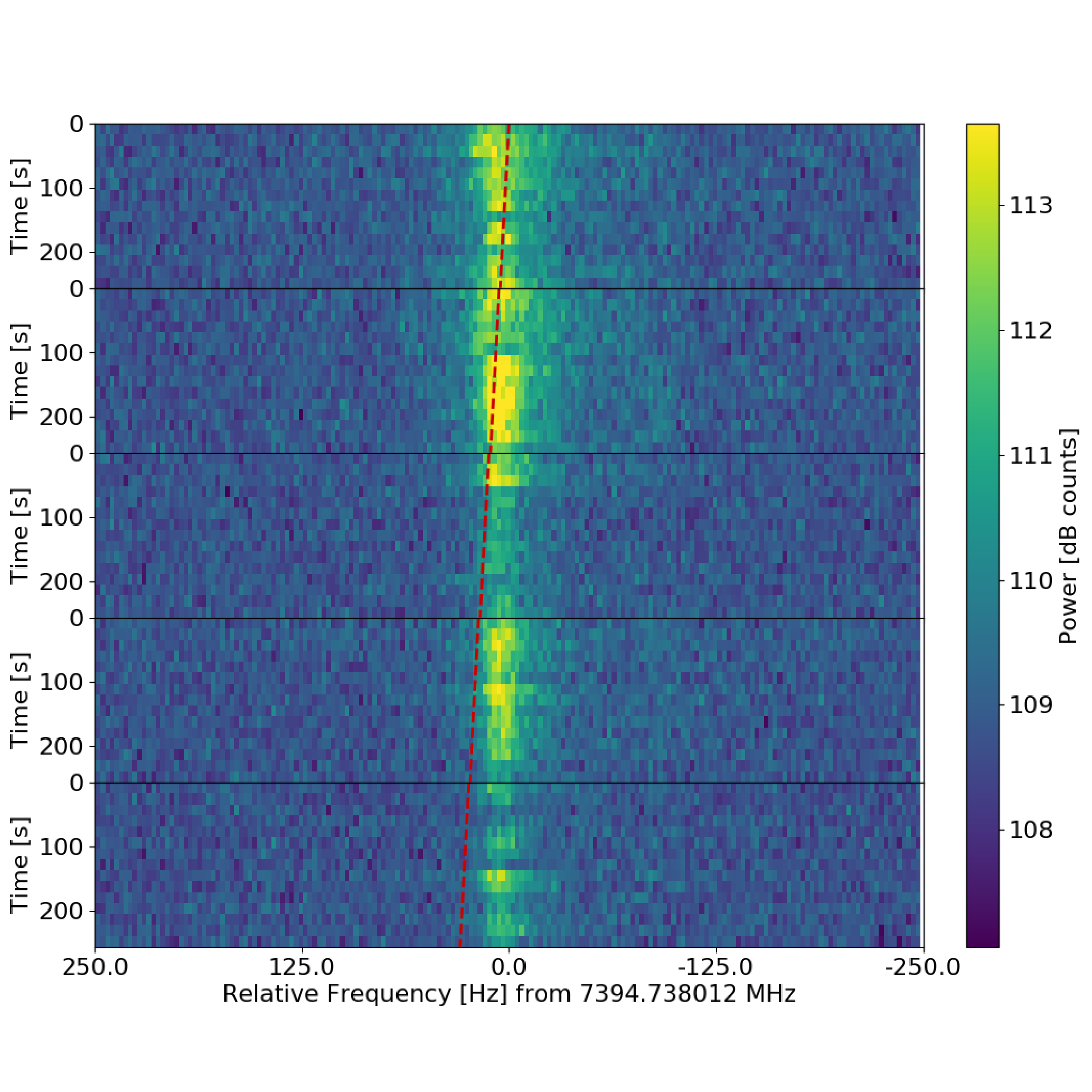}{0.3\textwidth}{(h)\label{fig:column}}
          \fig{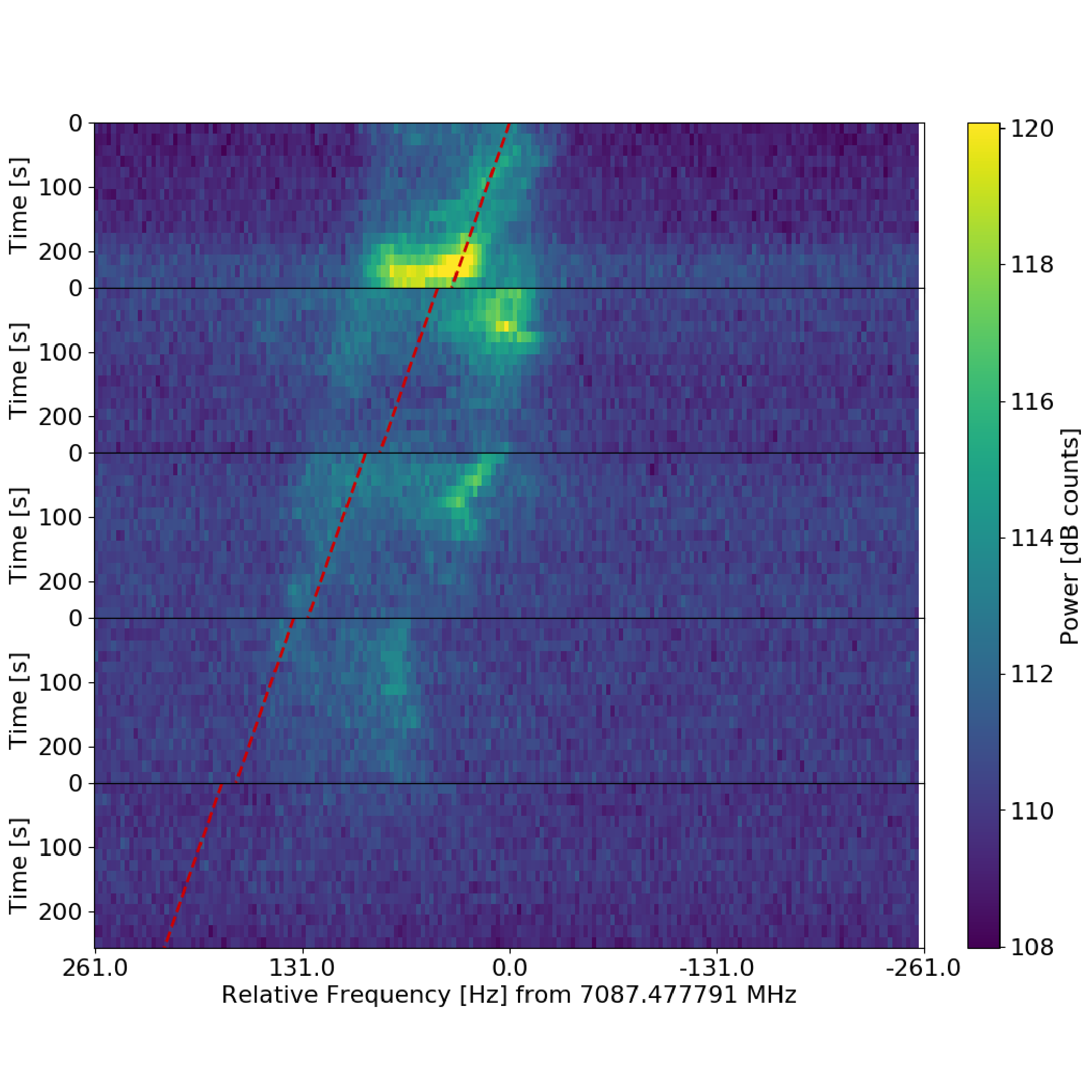}{0.3\textwidth}{(i)\label{fig:ghost}}}
        \caption{Nine dynamic spectra of events that are examples of the different morphologies of RFI recovered by the ETZ SETI search.  Regions in yellow have a strong intensity. The y-axis of each plot shows the progression of the observation through time from top to bottom. The horizontal axis shows offset in Hz from the measured frequency of the signal in the first panel. While all of these plots are visually interesting, and all of them passed at least one of the two filters, none of them are SETI candidates because the ``hit'' behaviour appears in both the ON and the OFF panels (alternating panels, top to bottom, in each dynamic spectrum). Each plot also shows the best-fit line that \textit{turboSETI} reported for the event (red dashed line).}
        \label{fig:rfi}
\end{figure}

The variety of RFI morphologies that we observed in this survey are shown in Figure \ref{fig:rfi}.

\textit{turboSETI} will often flag plots that with signals that appear in both the on-source and the off-source, and thus are clearly RFI by visual inspection. This can happen for many reasons. In some cases, the SNR is high enough to count as a hit in the on-source, but falls below the chosen SNR threshold in the off-source. This could be due to subtleties in the calculation of the signal (e.g. the signal often spans more than one 2.7 Hz channel, but only one channel is considered in the hit calculation) or subtleties in the noise (e.g. too much RFI in a coarse channel or a strong continuum source near the off-source position).

Some events in our sample, for example Figures \ref{fig:wave}d and \ref{fig:ghost}i, likely passed the \textit{turboSETI} thresholds because they have complex structures that drift in frequency, muddling the linear fit. In other cases, such as Figure \ref{fig:pulse}f, the RFI has an inherent long-timescale periodicity or amplitude variation that can match the observing cadence. Some cases of RFI, such as Figure \ref{fig:squiggle}e, have periodic or quasiperiodic frequency instabilities, suggesting either an alternating physical acceleration (e.g. of a poorly stabilized satellite) or an unstable oscillator. But others, such as the large morphology class of linear signals demonstrated by Figure \ref{fig:line}g are clearly RFI, probably carrier waves, and should ideally be rejected by the pipeline. However, regardless of their original source, which is outside the scope of this paper, all of these signals are clearly RFI because they appear in the on-source and the off-source and thus they will not be considered further.

\subsection{Potential Candidates}
\label{ssec: bestcand}

\begin{figure}[ht]
\gridline{\fig{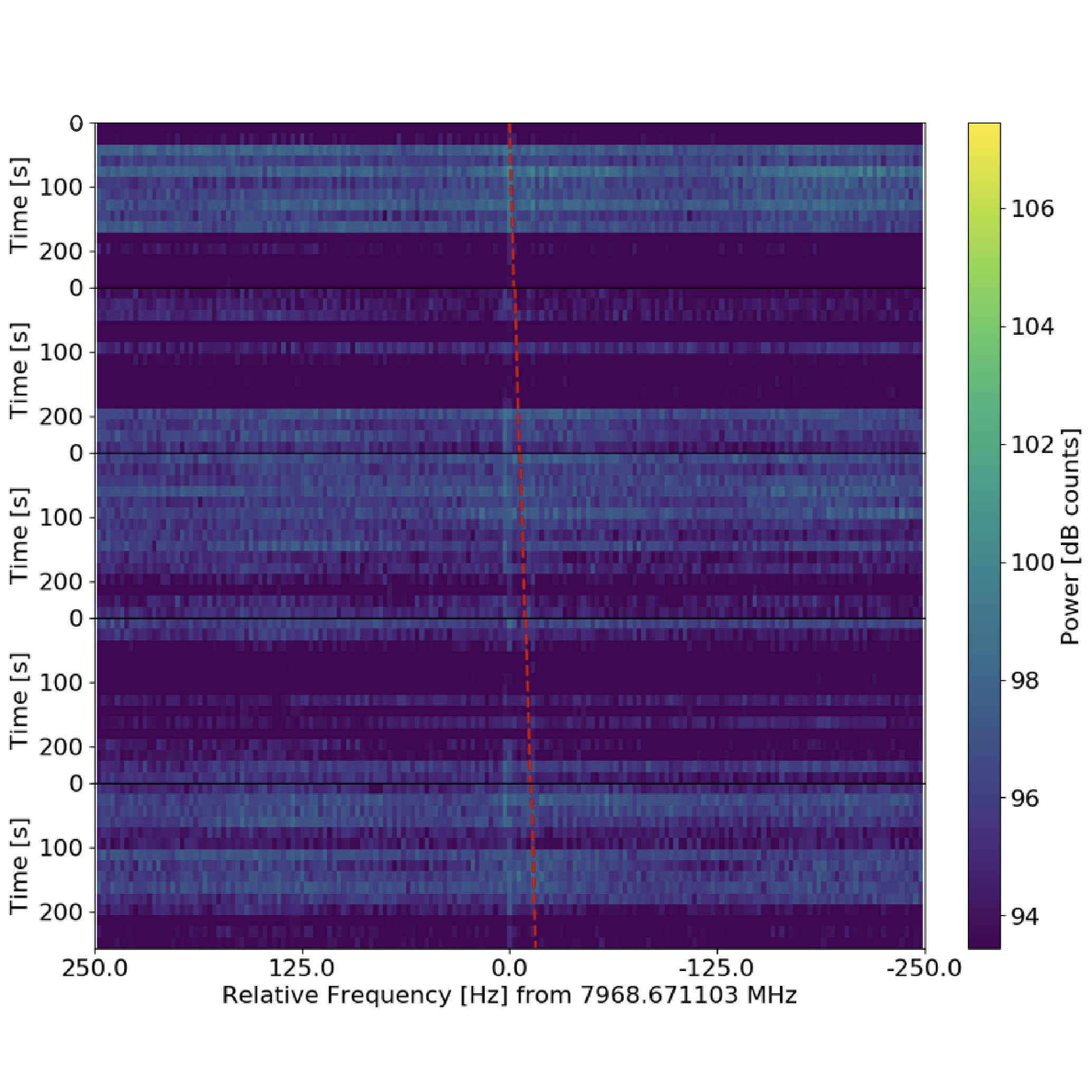}{0.32\textwidth}{(a)\label{fig:cand_one}}
          \fig{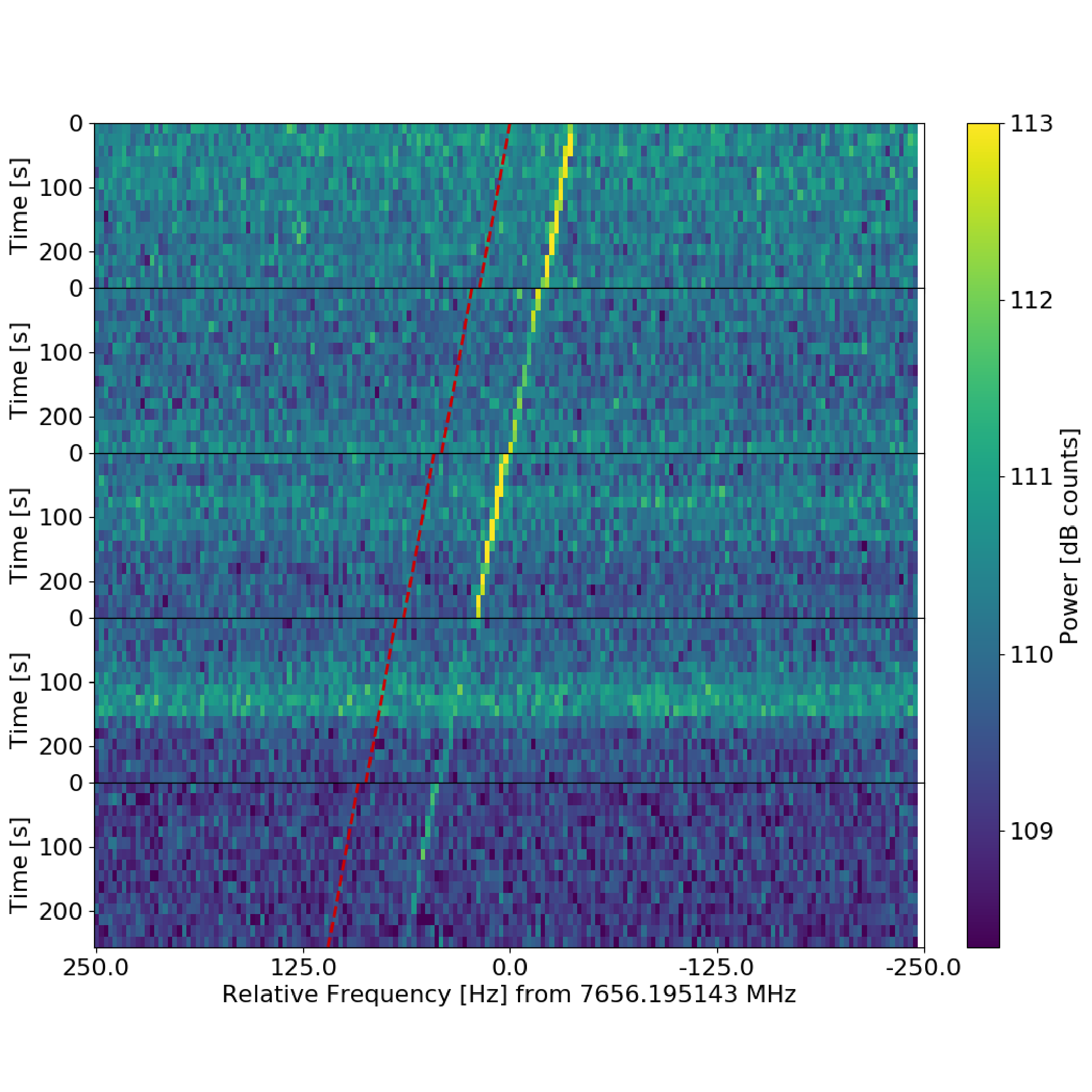}{0.32\textwidth}{(b)\label{fig:cand_two}}
          \fig{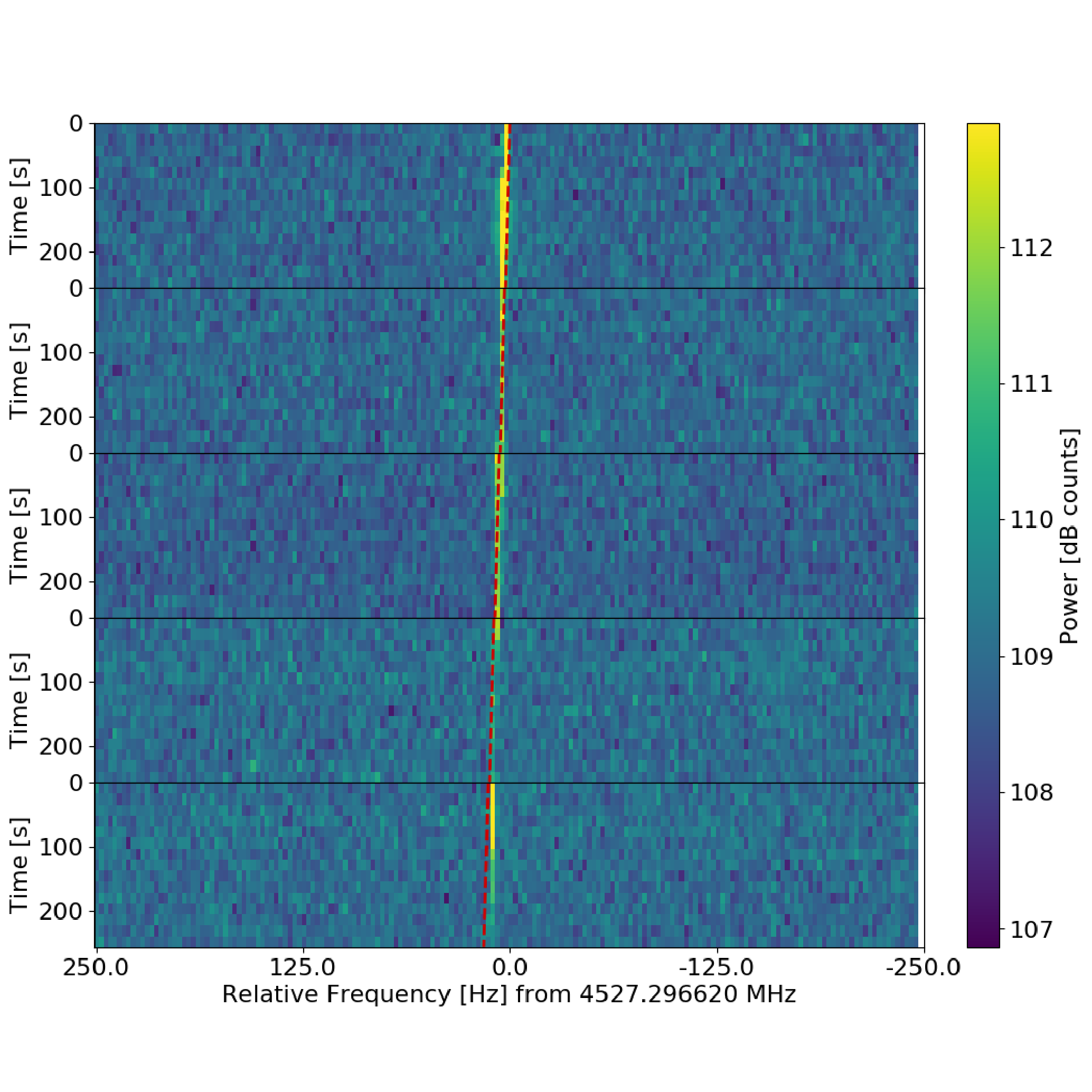}{0.32\textwidth}{(c)\label{fig:cand_three}}}
        \caption{Three output composite dynamic spectra for potential candidates detected in a) HIP 65642 b) HIP 96440 and c) HIP 88982. The five vertical panels are dynamic spectra from the ON-OFF-ON-OFF-ON observing sequence, progressing in time from top to bottom along the y-axis. The horizontal axis shows offset in Hz from the measured frequency of the signal in the first panel. Blue indicates areas of low intensity and yellow indicates areas of high intensity. All three plots are overlaid with a ``best-fit'' line (offset to the left for visibility in subfigure b). While all three narrowband signals vary in strength across the panels, they are all still visible in the OFF observations. These potential candidates are therefore radio frequency interference.}
        \label{fig:secondary_potential_candidates}
\end{figure}

\begin{figure}[ht]
    \centering
    \includegraphics[width=0.8\textwidth]{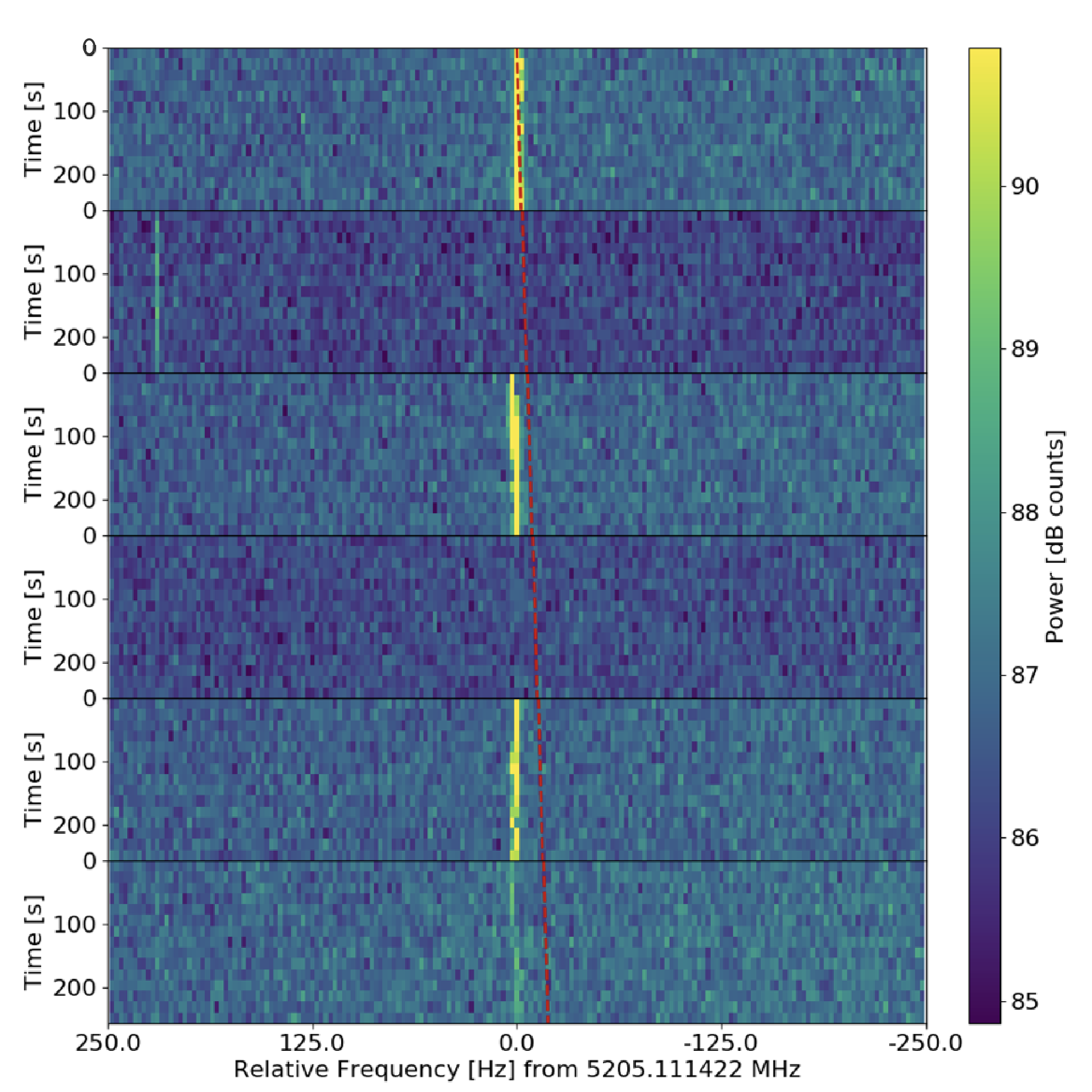}
    \caption{An output composite dynamic spectrum for a potential candidate detected in HIP 109656, overlaid with the best-fit line from \textit{turboSETI}. The six vertical panels are dynamic spectra from the ON-OFF-ON-OFF-ON-OFF observing sequence, progressing in time from top to bottom along the y-axis. The horizontal axis shows offset in Hz from the measured frequency of the signal in the first panel. The intensity colorbar represents power in decibel units of 10log$_{10}$(x) where x is the value in counts from the back-end. Blue indicates areas of low intensity and yellow indicates areas of high intensity. The best-fit line from \textit{turboSETI} has a drift rate of -0.01 Hzs$^{-1}$. This potential candidate is RFI because of the faint signal that appears in the final OFF panel, as well as its near zero drift rate. 0.01 Hzs$^{-1}$ is the lowest absolute drift rate that \textit{turboSETI} can currently measure and here the potential candidate has a large enough SNR in the first ON observation to trigger on -0.01 Hzs$^{-1}$ as well as on the natural zero drift rate of the signal. This causes the difference between the slope of the best-fit line and the apparent path of the signal. This signal also illustrates the importance of multiple ON-OFF pairs. Without the final OFF observation, this signal would not be identifiable as RFI and would be indistiguishable from an actual candidate. }
    \label{fig: cand_four}
\end{figure}

We found four groups of event plots that warranted closer inspection. These four groups consisted of 103 similar event plots that could be sorted into groups due to their repeated morphologies across frequencies with similar drift rates. Examples from three of the four sets are shown in Figure \ref{fig:secondary_potential_candidates}, while an example from the best potential candidate in our sample is shown in Figure \ref{fig: cand_four}. None of these potential candidates are consistent with the properties of a non-anthropogenic technosignature.

\begin{table}[ht]
    \centering
    \begin{tabular}{|c|c|c|c|c|c|c|}
    \hline
                    Target & Flagged By & \# of Events & Frequencies (MHz) & Drift Rates (Hz/s) & SNRs \\
                    \hline
    HIP 65642      &         AH          & 4         & 6631.73, 7968.67   & 0.00, 0.01       & 24, 37\\
    HIP 96440 and HIP 95417   & OND         & 12 and 5  & 7655.18--7656.41    & 0.05--0.07   & 10--20 \\
    HIP 88982      &         AH and OND  & 271        & 4506.04--4526.98    & 0.01        & 10--20 \\
    \hline
    LTT 88982      &  OND  & 94        & 5190.30--5213.93    & $\pm$0.01        & 10--180 \\
    \hline
    \end{tabular}
    \caption{The four potential candidate sets identified by the two filters. AH stands for ``All-Hit Filter'' and OND stands for ``Only Non-Zero Drift Filter''.}
    \label{tab: potential_candidates}
\end{table}

We investigated a set of events from HIP 65642, two sets of events from HIP 96440 and HIP 95417, and a set of events from HIP 88982. Examples of these events are shown in Figure \ref{fig:secondary_potential_candidates}, and their parameters are shown in the top three rows of Table \ref{tab: potential_candidates}. We combined the events from HIP 96440 and HIP 95417 because HIP 96440 was the off-source location for HIP 95417, and all of the flagged hits were similar in drift rate, morphology, and frequency. For HIP 88982, only a subset of 58 of the 271 events was originally plotted due to heavy RFI in this particular node. Note from Figure \ref{fig:secondary_potential_candidates}a that the narrowband component that \textit{turboSETI} flagged in HIP 65642 is part of a larger broadband quasi-periodic structure with a period similar to the observation cadence.

All of these potential candidates have intensity variations in the events between the ON and OFF observations, such that they passed the \textit{turboSETI} filters. However, we dismiss them as non-human technosignatures because:

\begin{enumerate}
    \item All of the signals appear at significant intensity in the OFF observations as well. Given GBT edge taper parameters from \citet{GBTSuppor2017} and radiation patterns for single-dish apertures, the pointing offset between the ON and OFF sources should have caused a drop in observed flux of >3 orders of magnitude if the source was localized on the sky. This drop was not observed. Interestingly, in the case of HIP 96440 and HIP 95417, the intensity changes depended on frequency such that some signals appeared to be events in HIP 96440 and some appeared to be events in HIP 95417, even though the signals were appearing at significant intensity in the other observation as well.
    \item 0.01 Hzs$^{-1}$ is the lowest absolute drift rate that \textit{turboSETI} can currently measure and can easily trigger on a zero drift rate signal if the signal has a large enough SNR. HIP 65642 and HIP 88982 therefore have drift rates consistent with zero drift.
\end{enumerate}

Taken all together, we dismiss these potential candidates and find them consistent with anthropogenic RFI.

We also identified a set of 19 events from HIP 109656 which we will discuss separately, as it was the most promising of the four sets. We investigated this set of events as a potential candidate after it was flagged by the Only Non-Zero Drift filter. One example from this set is shown in Figure \ref{fig: cand_four} and its parameters are shown in the last row of Table \ref{tab: potential_candidates}. These 19 events were determined to be a subset of 94 events on the same target with similar frequencies and drift rates, but only the subset of 19 were originally plotted due to heavy RFI in this particular observing session. These 94 total events had drift rates of $\pm$0.01 Hzs$^{-1}$, a varied range of signal-to-noise ratios ranging from 10 to 180, and a range of frequencies of 5190.30--5213.93 MHz. It should be noted that seeing repeated signals across multiple frequencies in a single target does not argue for or against a potential candidate's validity; both signals from extraterrestrial intelligence and anthropogenic RFI could both potentially be present at multiple frequencies simultaneously.

As can be seen in Figure \ref{fig: cand_four}, the signal appears completely absent in the first and second OFF observations, and only appears weakly in the final OFF observation. This weak signal is actually reported as an event a few times in the final OFF, at an SNR of around 12. The distance between HIP 109656 and its off-source, HIP 111332, is $\sim$ 5.5$^\circ$ --- at C-band with the GBT, this is equivalent to $\sim $170 beamwidths. Even at a signal-to-noise ratio of 180, in the brightest case, we would not expect this signal to also be present in the final off-source observation at an SNR of 12 if it was actually coming from HIP 109656's position on the sky. Without the final OFF observation this potential candidate would be indistiguishable from an actual candidate, illustrating the importance of having multiple ON-OFF pairs for RFI rejection. 

In addition, every single detection had the minimum detectable drift rate of $\pm$0.01 Hzs$^{-1}$, one drift rate bin away from the zero-drift rate signature of Earth-based RFI. This can occur when a zero-drift rate signal has a strong enough SNR, and it can be seen in Figure \ref{fig: cand_four} that the signal does not actually deviate from its original frequency in the first ON panel. To elaborate: the drift rate expected for a true extraterrestrial signal in the barycentric frame is non-zero. The terrestrial drift rate contribution to an observation primarily comes from Earth's rotation, meaning that each terrestrial observatory has a different ``zero'' drift rate for each time and target. It would thus be extremely unlikely for a true signal to happen to align with the GBT's zero drift rate during an observation.

We reject this as an extraterrestrial source primarily because it is present in an off pointing, and secondarily because the drift rate is very close to zero, which has no astrophysical significance but is consistent with most known RFI.

\newpage
\section{Discussion}
\label{sec: discussion}

\citet{enriquez2017breakthrough} defines a Continuous Wave Transmitter Figure of Merit (CWTFM) for SETI searches to quantify the merits of various searches. The CWTFM equation is shown in Equation \ref{eq: cwtfm}.

\begin{equation}
    \textrm{CWTFM} = \zeta_{AO} \frac{\textrm{EIRP}_{\textrm{min}} \nu_{\textrm{mid}}}{N_\ast \Delta \nu_{\textrm{tot}}}
    \label{eq: cwtfm}
\end{equation}

$\zeta_{AO}$ is a normalization factor of $5 \times 10^{-11} \textrm{W}^{-1}$, $\textrm{EIRP}_{\textrm{min}}$ is the smallest EIRP detectable for \textit{all} stars in the sample (i.e. the largest value of $\textrm{EIRP}_{\textrm{min}}$ in Table \ref{tab: target_details}), $\nu_{\textrm{mid}}$ is the central frequency of the observation, $N_\ast$ is the number of stars in the survey, and $\Delta \nu_{\textrm{tot}}$ is the total bandwidth of the observation. For this survey, the CWTFM = 87.7. For a survey with only 20 targets, this is quite impressive -- because all of the targets in this survey are nearby, we are sensitive to transmitters with very low EIRPs compared to other searches. The value for this work is compared with the CWTFMs for other searches in Table \ref{tab:cwtfm}.

\begin{table}[ht]
    \centering
    \begin{tabular}{|c|c|}
    \hline
         Search                             & CWTFM \\
    \hline
         \citet{enriquez2017breakthrough}   & 0.85  \\
         Project Phoenix                    & 49    \\
        This Work                          & 87.7  \\
        \citet{gray2017vla}                   & 136.3 \\
        \citet{siemion20131}               & 1878  \\
         \citet{harp2016seti}               & 2970  \\
    \hline
    \end{tabular}
    \caption{Continuous Wave Transmitter Figures of Merit (CWTFM) for this work compared to other searches. Values for the other searches were taken from \citet{enriquez2017breakthrough}. This work scores relatively well because all of the targets are nearby, allowing us to search for transmitters with very low EIRPs.}
    \label{tab:cwtfm}
\end{table}

\citet{Sheikh2019a} has recently suggested a drift rate choice of 200 nHz for radio SETI searches, or approximately 10 times the value chosen for the highest frequency observations in this survey. While the value chosen for this work was considered to be abnormally high in the context of past searches, future work should try to expand the range of drift rates searched in order to account for reasonable physical accelerations from all currently observed exoplanetary systems \citep{Sheikh2019a}. In addition, in future searches, the hit identification step should be performed by searching a \textit{normalized} drift range, which will require a dynamically shifting Hzs$^{-1}$ upper limit depending on the rest frequency of each channel in the observation. The physical significance of a 10 Hzs$^{-1}$ drift rate will change based on the observing frequency, which is why the normalized value (divided by the observing frequency) should be preferred.

This study provides a concrete example of how observations in C-band (4--8 GHz) contain much less RFI than other frequency ranges. The hit rate density for this search was $8.86 \times 10^{-9}$ hits per hour per Hz. From \citet{Price2019}, comparable Breakthrough Listen observations at Green Bank in L-band (1.10--1.90 GHz) and S-Band (1.80--2.80 GHz) had hit rate densities of $1.11 \times 10^{-4}$ and $2.00 \times 10^{-5}$ hits per hour per Hz, respectively. Thus, the RFI environment at the GBT is $10^4$--$10^5$ times less crowded in C-band than in L-band or S-band. 

This survey also only observed a total of 20 target stars in the rETZ and was restricted to spectral types G and K as per Table 1 in \citet{heller2016search}. This study has not exhausted the possibility of a SETI signal in the rETZ, but has only begun a new category of searches. Combining this particular region of the sky with concepts such as Schelling points \citep{wright2018recommendations} could begin to search an interesting and heretofore unexplored parameter space. One potential project would be to observe the target at a particular Schelling time deducible by both the transmitter and the receiver. One example is the anti-Solar point: the geometry when the Earth would appear to be at the center of its transit across the Sun from the target's point of view.

Stars in this survey were chosen to be nearby, so they have correspondingly high proper motions. As noted in \citet{heller2016search}, the stars in the ETZ are not static. Stellar proper motions combined with the variation of the ecliptic will cause stars to enter and leave the ETZ over timescales of thousands of years. For this reason, stars that are within approximately 1$^\circ$ of the ecliptic should also be considered in future searches of the ETZ \citep{heller2016search}.

The ETZ method can also be used in conjunction with the idea of ``synchrosignals'' \citep{makovetskii1977nova} or temporal Schelling Points \citep{wright2018recommendations}. For a given object within the ETZ, there is a ``special'' time, the Earth's transit from their vantage point \citep[or, from Earth, ``opposition'' as in][]{corbet2003synchronized}, that is known to both the observer on Earth and the hypothetical transmitter on the object. This provides a lower-energy alternative to a continuous signal and has been suggested for the ecliptic specifically in \citet{filippova1992bioastronomy}. We did not temporally align our observations with this transit time, but further studies of the ETZ could take advantage of this combined method.

\section{Conclusions}
\label{sec: conclusions}

In this study, the Breakthrough Listen project performed the first SETI search of the restricted Earth Transit Zone (rETZ) --- the region of the sky around the ecliptic from which an observer would be able to see the Earth transit the Sun. In Section \ref{sec: targetselection}, we selected 20 nearby targets from a list suggested by \citet{heller2016search}. In the parlance of \citet{wright2018much}, we performed this targeted search in order to set an upper limit on detection from these systems in a frequency range of 3.95--8.00 GHz, with a sensitivity to signals above 1.01 $\times$ 10$^{-25}$ W/m$^2$, looking at both polarizations, with repetition rates below 288 seconds, of narrowband modulation (Section \ref{sec: searchparameters}). We observed these targets with the Green Bank Telescope for three observations of 288 seconds each, interspersed with off-source observations of a nearby star for later RFI rejection (Section \ref{sec: observations}).

Using methods from the software packages \textit{blimpy} and \textit{turboSETI}, we searched for strong (signal-to-noise ratio > 10) narrowband hits drifting with a range of drift rates from -20 to 20 Hzs$^{-1}$ in the data from the 20 rETZ targets as well as 10 off-source stars chosen solely for their proximity to one of the rETZ targets (Section \ref{sec: dataanalysis}). We found 425168 total hits in these 30 stars for our observations. These hits were then grouped into 2986 related ``events'' which were identified by two search pipelines with different search parameters. Of those events, after inspection of each event's dynamic spectrum by eye, we selected 103 event plots for further investigation (Section \ref{sec: results}). We then grouped these 103 events into four sets based on similar morphology and appearance in the same target, and examined each set more closely. Upon closer inspection, all four sets were clearly radio frequency interference.

We thus find no evidence for 100\%-duty cycle transmitters from 3.95--8.00 GHz directed at Earth with a power output equal to or greater than that of 88\% of the Arecibo radio transmitter in any of the star systems observed. In most systems we were sensitive to signals far less powerful than 88\% of Arecibo, but as noted in \citet{enriquez2017breakthrough}, using the most distant target to get an EIRP sensitivity that is true for \textit{all} targets biases the reporting of the result based on the most distant star. 

The Gaia Data Release 2 \citep{brown2018gaia} catalog reports 252 systems within the rETZ within 150 pc of the Sun. Assuming that the existence of extraterrestrial intelligences is independent of their host system's distance from Earth, our results suggest that at least $8\%$ of the systems in the rETZ within 150 pc do not possess the types of transmitters described above. While Gaia has a high completeness to G and K dwarfs such as those used in this study, M-dwarfs, especially at the far edge of the sample, are going to be underreported; the number of stars within the rETZ and within 150 pc is thus greater than the 252 number reported above. As discussed in Section \ref{sec: discussion}, this is only the first cursory attempt at studying this particular region. There is much parameter space still to search in the ETZ so future studies are needed to constrain the prevalence of extraterrestrial intelligences. 

\software{astropy \citep{robitaille2013astropy, pricewhelan2018astropy}, 
blimpy \citep{Price2019blimpy},
turboSETI \citep{enriquez2017breakthrough}}
sigproc \citep{lorimer2011sigproc}
numpy \citep{developers2013numpy}
matplotlib \citep{hunter2007matplotlib}

\acknowledgments

Breakthrough Listen is managed by the Breakthrough Initiatives, sponsored by the Breakthrough Prize Foundation. The Green Bank Observatory is a facility of the National Science Foundation, operated under cooperative agreement by Associated Universities, Inc. This research has made use of the Exoplanet Orbit Database and the Exoplanet Data Explorer at exoplanets.org, as well as the SIMBAD database, operated at CDS, Strasbourg, France. SZS would like to acknowledge the invaluable assistance of the staff and students of the Berkeley SETI Research Center and the Green Bank Observatory, who made this project possible. SZS would also like to thank Jason Wright for general advising, Shara Hussain for assistance with RFI classification, and Zayna Sheikh for assistance with data visualization.

\bibliographystyle{aasjournal}
\bibliography{references.bib}

\begin{thebibliography}{}
\expandafter\ifx\csname natexlab\endcsname\relax\def\natexlab#1{#1}\fi

\bibitem[{Arnold(2005)}]{Arnold2005}
Arnold, L. F.~A. 2005, The Astrophysical Journal, 627, 534

\bibitem[{{Bidelman}(1985)}]{bidelman1985astronomical}
{Bidelman}, W.~P. 1985, \apjs, 59, 197

\bibitem[{Borucki {et~al.}(2008)Borucki, Koch, Batalha, Caldwell,
  Christensen-Dalsgaard, Cochran, Dunham, Gautier, Geary, Gilliland, Jenkins,
  Kjeldsen, Lissauer, \& Rowe}]{Borucki2008}
Borucki, W., Koch, D., Batalha, N., {et~al.} 2008, Proceedings of the
  International Astronomical Union, 4, 289

\bibitem[{Bracewell(1960)}]{Bracewell1960}
Bracewell, R.~N. 1960, Nature, 186, 670

\bibitem[{Breckenridge \& Kukolich(2002)}]{Breckenridge2002}
Breckenridge, S.~M., \& Kukolich, S.~G. 2002, The Astrophysical Journal, 438,
  504

\bibitem[{Brown {et~al.}(2018)Brown, Vallenari, Prusti, De~Bruijne, Babusiaux,
  Bailer-Jones, Biermann, Evans, Eyer, Jansen, {et~al.}}]{brown2018gaia}
Brown, A., Vallenari, A., Prusti, T., {et~al.} 2018, Astronomy \& astrophysics,
  616, A1

\bibitem[{{Cannon} \& {Pickering}(1993)}]{cannon1993vizier}
{Cannon}, A.~J., \& {Pickering}, E.~C. 1993, VizieR Online Data Catalog,
  III/135A

\bibitem[{Castellano {et~al.}(2004)Castellano, Doyle, \&
  McIntosh}]{castellano2004visibility}
Castellano, T., Doyle, L., \& McIntosh, D. 2004, in Symposium-International
  Astronomical Union, Vol. 202, Cambridge University Press, 445--447

\bibitem[{Cocconi \& Morrison(1959)}]{Cocconi1959}
Cocconi, G., \& Morrison, P. 1959, Nature, 184, 844

\bibitem[{Conn~Henry {et~al.}(2008)Conn~Henry, Kilston, \&
  Shostak}]{conn2008seti}
Conn~Henry, R., Kilston, S., \& Shostak, S. 2008, in Bulletin of the American
  Astronomical Society, Vol.~40, 194

\bibitem[{Corbet(2003)}]{corbet2003synchronized}
Corbet, R.~H. 2003, Astrobiology, 3, 305

\bibitem[{Cordes {et~al.}(1997)Cordes, Lazio, \& Sagan}]{Cordes1997}
Cordes, J.~M., Lazio, T. J.~W., \& Sagan, C. 1997, The Astrophysical Journal,
  487, 782

\bibitem[{Crossfield {et~al.}(2016)Crossfield, Ciardi, Petigura, Sinukoff,
  Schlieder, Howard, Beichman, Isaacson, Dressing, Christiansen,
  {et~al.}}]{crossfield2016197}
Crossfield, I.~J., Ciardi, D.~R., Petigura, E.~A., {et~al.} 2016, The
  Astrophysical Journal Supplement Series, 226, 7

\bibitem[{Developers(2013)}]{developers2013numpy}
Developers, N. 2013, NumPy Numpy. Scipy Developers, 31

\bibitem[{Dommanget \& Nys(2002)}]{dommanget2002vizier}
Dommanget, J., \& Nys, O. 2002, VizieR Online Data Catalog, 1274

\bibitem[{Drake(1961)}]{Drake1961}
Drake, F.~D. 1961, Physics Today, 14, 40

\bibitem[{Dressing \& Charbonneau(2015)}]{dressing2015occurrence}
Dressing, C.~D., \& Charbonneau, D. 2015, The Astrophysical Journal, 807, 45

\bibitem[{{Dressing} {et~al.}(2017){Dressing}, {Newton}, {Schlieder},
  {Charbonneau}, {Knutson}, {Vanderburg}, \&
  {Sinukoff}}]{dressing2017characterizing}
{Dressing}, C.~D., {Newton}, E.~R., {Schlieder}, J.~E., {et~al.} 2017, \apj,
  836, 167

\bibitem[{Dyson(1960)}]{Dyson1960}
Dyson, F.~J. 1960, Science, 131, 1667

\bibitem[{Dyson(1966)}]{dyson1966search}
---. 1966, Perspectives in modern physics, 641

\bibitem[{Enriquez {et~al.}(2017)Enriquez, Siemion, Foster, Gajjar, Hellbourg,
  Hickish, Isaacson, Price, Croft, DeBoer, {et~al.}}]{enriquez2017breakthrough}
Enriquez, J.~E., Siemion, A., Foster, G., {et~al.} 2017, The Astrophysical
  Journal, 849, 104

\bibitem[{{Fabricius} {et~al.}(2002){Fabricius}, {H{\o}g}, {Makarov}, {Mason},
  {Wycoff}, \& {Urban}}]{fabricius2002tycho}
{Fabricius}, C., {H{\o}g}, E., {Makarov}, V.~V., {et~al.} 2002, \aap, 384, 180

\bibitem[{Filippova(1990)}]{filippova1990list}
Filippova, L. 1990, Astronomicheskij Tsirkulyar, 1544, 37

\bibitem[{{Filippova} \& {Strelnitskij}(1988)}]{filippova1988ecliptic}
{Filippova}, L.~N., \& {Strelnitskij}, V.~S. 1988, Astronomicheskij Tsirkulyar,
  1531, 31

\bibitem[{Filippova(1992)}]{filippova1992bioastronomy}
Filippova, Kardashev, L.~S. 1992, in Bioastronomy: The Search for
  Extraterrestrial Life--The Exploration Broadens, ed. J.~Heidmann \& M.~J.
  Klein, Vol. 390 (Berlin Springer Verlag), 254--258

\bibitem[{Gardner {et~al.}(2006)Gardner, Mather, Clampin, Doyon, Greenhouse,
  Hammel, Hutchings, Jakobsen, Lilly, Long, Lunine, Mccaughrean, Mountain,
  Nella, Rieke, Rieke, Rix, Smith, Sonneborn, Stiavelli, Stockman, Windhorst,
  \& Wright}]{Gardner2006}
Gardner, J.~P., Mather, J.~C., Clampin, M., {et~al.} 2006, Space Science
  Reviews, 123, 485

\bibitem[{{GBT Support Staff}(2017)}]{GBTSuppor2017}
{GBT Support Staff}. 2017, {Green Bank Observatory Proposer's Guide for the
  Green Bank Telescope}, Tech. rep., Green Bank Observatory

\bibitem[{Gray \& Mooley(2017)}]{gray2017vla}
Gray, R.~H., \& Mooley, K. 2017, The Astronomical Journal, 153, 110

\bibitem[{{Gray} {et~al.}(2006){Gray}, {Corbally}, {Garrison}, {McFadden},
  {Bubar}, {McGahee}, {O'Donoghue}, \& {Knox}}]{Gray2006contributions}
{Gray}, R.~O., {Corbally}, C.~J., {Garrison}, R.~F., {et~al.} 2006, \aj, 132,
  161

\bibitem[{Harp {et~al.}(2016)Harp, Richards, Tarter, Dreher, Jordan, Shostak,
  Smolek, Kilsdonk, Wilcox, Wimberly, {et~al.}}]{harp2016seti}
Harp, G., Richards, J., Tarter, J.~C., {et~al.} 2016, arXiv preprint
  arXiv:1607.04207

\bibitem[{{Heckmann}(1975)}]{heckmann1975agk}
{Heckmann}, O. 1975, {AGK 3. Star catalogue of positions and proper motions
  north of -2.5 deg. declination} (n.p.)

\bibitem[{Heller \& Pudritz(2016)}]{heller2016search}
Heller, R., \& Pudritz, R.~E. 2016, Astrobiology, 16, 259

\bibitem[{{H{\o}g} {et~al.}(2000){H{\o}g}, {Fabricius}, {Makarov}, {Urban},
  {Corbin}, {Wycoff}, {Bastian}, {Schwekendiek}, \& {Wicenec}}]{hog2000tycho}
{H{\o}g}, E., {Fabricius}, C., {Makarov}, V.~V., {et~al.} 2000, \aap, 355, L27

\bibitem[{{Houk} \& {Smith-Moore}(1988)}]{houk1988michigan}
{Houk}, N., \& {Smith-Moore}, M. 1988, {Michigan Catalogue of Two-dimensional
  Spectral Types for the HD Stars. Volume 4, Declinations -26.0 to -12.0.},
  Vol.~4 (n.p)

\bibitem[{{Houk} \& {Swift}(1999)}]{houk1999michigan}
{Houk}, N., \& {Swift}, C. 1999, Michigan Spectral Survey, 5, 0

\bibitem[{Howard {et~al.}(2004)Howard, Horowitz, Wilkinson, Coldwell, Groth,
  Jarosik, Latham, Stefanik, {Willman, Jr.}, Wolff, \& Zajac}]{Howard2004}
Howard, A.~W., Horowitz, P., Wilkinson, D.~T., {et~al.} 2004, The Astrophysical
  Journal, 613, 1270

\bibitem[{Hunter(2007)}]{hunter2007matplotlib}
Hunter, J.~D. 2007, Computing in science \& engineering, 9, 90

\bibitem[{Isaacson {et~al.}(2017)Isaacson, Siemion, Marcy, Lebofsky, Price,
  MacMahon, Croft, DeBoer, Hickish, Werthimer,
  {et~al.}}]{isaacson2017breakthrough}
Isaacson, H., Siemion, A.~P., Marcy, G.~W., {et~al.} 2017, Publications of the
  Astronomical Society of the Pacific, 129, 054501

\bibitem[{Jones {et~al.}(2011)Jones, Jenkins, Rojo, \& Melo}]{jones2011study}
Jones, M., Jenkins, J., Rojo, P., \& Melo, C. 2011, Astronomy \& Astrophysics,
  536, A71

\bibitem[{Krissansen-Totton {et~al.}(2018)Krissansen-Totton, Garland, Irwin, \&
  Catling}]{Krissansen-Totton2018}
Krissansen-Totton, J., Garland, R., Irwin, P., \& Catling, D.~C. 2018, The
  Astronomical Journal, 156, 114

\bibitem[{{Kunder} {et~al.}(2017){Kunder}, {Kordopatis}, {Steinmetz},
  {Zwitter}, {McMillan}, {Casagrande}, {Enke}, {Wojno}, {Valentini},
  {Chiappini}, {Matijevi{\v{c}}}, {Siviero}, {de Laverny}, {Recio-Blanco},
  {Bijaoui}, {Wyse}, {Binney}, {Grebel}, {Helmi}, {Jofre}, {Antoja}, {Gilmore},
  {Siebert}, {Famaey}, {Bienaym{\'e}}, {Gibson}, {Freeman}, {Navarro},
  {Munari}, {Seabroke}, {Anguiano}, {{\v{Z}}erjal}, {Minchev}, {Reid},
  {Bland-Hawthorn}, {Kos}, {Sharma}, {Watson}, {Parker}, {Scholz}, {Burton},
  {Cass}, {Hartley}, {Fiegert}, {Stupar}, {Ritter}, {Hawkins}, {Gerhard},
  {Chaplin}, {Davies}, {Elsworth}, {Lund}, {Miglio}, \&
  {Mosser}}]{kunder2017rave}
{Kunder}, A., {Kordopatis}, G., {Steinmetz}, M., {et~al.} 2017, \aj, 153, 75

\bibitem[{Lebofsky(2020)}]{lebofsky2020datarelease}
Lebofsky, M. 2020, in prep.

\bibitem[{{Lebofsky} {et~al.}(2019){Lebofsky}, {Croft}, {Siemion}, {Price},
  {Enriquez}, {Isaacson}, {MacMahon}, {Anderson}, {Brzycki}, {Cobb}, {Czech},
  {DeBoer}, {DeMarines}, {Drew}, {Foster}, {Gajjar}, {Gizani}, {Hellbourg},
  {Korpela}, {Lacki}, {Sheikh}, {Werthimer}, {Worden}, {Yu}, \&
  {Zhang}}]{Lebofsky2019}
{Lebofsky}, M., {Croft}, S., {Siemion}, A. P.~V., {et~al.} 2019, \pasp, 131,
  124505

\bibitem[{Lin {et~al.}(2014)Lin, Abad, \& Loeb}]{Lin2014}
Lin, H.~W., Abad, G.~G., \& Loeb, A. 2014, Astrophysical Journal Letters, 792,
  doi:10.1088/2041-8205/792/1/L7

\bibitem[{Lockwood \& Thompson(2009)}]{lockwood2009seasonal}
Lockwood, G., \& Thompson, D. 2009, Icarus, 200, 616

\bibitem[{Lorimer(2011)}]{lorimer2011sigproc}
Lorimer, D. 2011, Astrophysics Source Code Library

\bibitem[{{MacMahon} {et~al.}(2018){MacMahon}, {Price}, {Lebofsky}, {Siemion},
  {Croft}, {DeBoer}, {Enriquez}, {Gajjar}, {Hellbourg}, {Isaacson},
  {Werthimer}, {Abdurashidova}, {Bloss}, {Brandt}, {Creager}, {Ford}, {Lynch},
  {Maddalena}, {McCullough}, {Ray}, {Whitehead}, \&
  {Woody}}]{macmahon2017recorder}
{MacMahon}, D. H.~E., {Price}, D.~C., {Lebofsky}, M., {et~al.} 2018, \pasp,
  130, 044502

\bibitem[{Makovetskii(1977)}]{makovetskii1977nova}
Makovetskii, P. 1977, Soviet Astronomy, 21, 251

\bibitem[{Mayor \& Queloz(1995)}]{Mayor1995}
Mayor, M., \& Queloz, D. 1995, Nature, 378, 355

\bibitem[{{Nesterov} {et~al.}(1995){Nesterov}, {Kuzmin}, {Ashimbaeva},
  {Volchkov}, {R{\"o}ser}, \& {Bastian}}]{nesterov1995henry}
{Nesterov}, V.~V., {Kuzmin}, A.~V., {Ashimbaeva}, N.~T., {et~al.} 1995, \aaps,
  110, 367

\bibitem[{Nussinov(2009)}]{nussinov2009some}
Nussinov, S. 2009, arXiv preprint arXiv:0903.1628

\bibitem[{{Oja}(1987)}]{oja1987photometry}
{Oja}, T. 1987, \aaps, 71, 561

\bibitem[{Petigura {et~al.}(2013)Petigura, Howard, \&
  Marcy}]{petigura2013prevalence}
Petigura, E.~A., Howard, A.~W., \& Marcy, G.~W. 2013, Proceedings of the
  National Academy of Sciences, 110, 19273

\bibitem[{Pourbaix {et~al.}(2004)Pourbaix, Tokovinin, Batten, Fekel, Hartkopf,
  Levato, Morrell, Torres, \& Udry}]{pourbaix2004s}
Pourbaix, D., Tokovinin, A.~A., Batten, A.~H., {et~al.} 2004, Astronomy \&
  Astrophysics, 424, 727

\bibitem[{{Prestage} {et~al.}(2015){Prestage}, {Bloss}, {Brandt}, {Chen},
  {Creager}, {Demorest}, {Ford}, {Jones}, {Kepley}, {Kobelski}, {Marganian},
  {Mello}, {McMahon}, {McCullough}, {Ray}, {Roshi}, {Werthimer}, \&
  {Whitehead}}]{Prestage2015}
{Prestage}, R.~M., {Bloss}, M., {Brandt}, J., {et~al.} 2015, in 2015 URSI-USNC
  Radio Science Meeting, 4

\bibitem[{{Price} {et~al.}(2019){Price}, {Enriquez}, {Chen}, \&
  {Siebert}}]{Price2019blimpy}
{Price}, D., {Enriquez}, J., {Chen}, Y., \& {Siebert}, M. 2019, The Journal of
  Open Source Software, 4, 1554

\bibitem[{Price {et~al.}(2020)Price, Enriquez, Brzycki, Croft, Czech, DeBoer,
  DeMarines, Foster, Gajjar, Gizani, {et~al.}}]{Price2019}
Price, D.~C., Enriquez, J.~E., Brzycki, B., {et~al.} 2020, The Astronomical
  Journal, 159, 86

\bibitem[{Price-Whelan {et~al.}(2018)Price-Whelan, Sip{\H{o}}cz, G{\"u}nther,
  Lim, Crawford, Conseil, Shupe, Craig, Dencheva, Ginsburg,
  {et~al.}}]{pricewhelan2018astropy}
Price-Whelan, A.~M., Sip{\H{o}}cz, B., G{\"u}nther, H., {et~al.} 2018, The
  Astronomical Journal, 156, 123

\bibitem[{Prusti {et~al.}(2016)Prusti, De~Bruijne, Brown, Vallenari, Babusiaux,
  Bailer-Jones, Bastian, Biermann, Evans, Eyer, {et~al.}}]{prusti2016gaia}
Prusti, T., De~Bruijne, J., Brown, A.~G., {et~al.} 2016, Astronomy \&
  Astrophysics, 595, A1

\bibitem[{Ransom(2011)}]{Ransom2011}
Ransom, S. 2011, Astrophysics Source Code Library, record ascl:1107.017

\bibitem[{Ricker {et~al.}(2014)Ricker, Winn, \& Vanderspek}]{Ricker2014}
Ricker, G., Winn, J., \& Vanderspek, R. 2014, Journal of Astronomical
  Telescopes, Instruments, and Systems, 1, 014003

\bibitem[{Robitaille {et~al.}(2013)Robitaille, Tollerud, Greenfield,
  Droettboom, Bray, Aldcroft, Davis, Ginsburg, Price-Whelan, Kerzendorf,
  {et~al.}}]{robitaille2013astropy}
Robitaille, T.~P., Tollerud, E.~J., Greenfield, P., {et~al.} 2013, Astronomy \&
  Astrophysics, 558, A33

\bibitem[{Schwartz \& Townes(1961)}]{Schwartz1961}
Schwartz, R.~N., \& Townes, C.~H. 1961, Nature, 190, 205

\bibitem[{Sheikh(2019)}]{Sheikh2019b}
Sheikh, S.~Z. 2019, International Journal of Astrobiology, 1

\bibitem[{{Sheikh} {et~al.}(2019){Sheikh}, {Wright}, {Siemion}, \&
  {Enriquez}}]{Sheikh2019a}
{Sheikh}, S.~Z., {Wright}, J.~T., {Siemion}, A., \& {Enriquez}, J.~E. 2019,
  \apj, 884, 14

\bibitem[{Shostak \& Villard(2004)}]{shostak_villard_2004}
Shostak, S., \& Villard, R. 2004, Symposium - International Astronomical Union,
  213, 409–414

\bibitem[{Siemion {et~al.}(2013)Siemion, Demorest, Korpela, Maddalena,
  Werthimer, Cobb, Howard, Langston, Lebofsky, Marcy, {et~al.}}]{siemion20131}
Siemion, A.~P., Demorest, P., Korpela, E., {et~al.} 2013, The Astrophysical
  Journal, 767, 94

\bibitem[{{Stephenson}(1986)}]{stephenson1986dwarf}
{Stephenson}, C.~B. 1986, \aj, 92, 139

\bibitem[{Tarter(2006)}]{Tarter2006}
Tarter, J.~C. 2006, Proceedings of the International Astronomical Union, 2, 14

\bibitem[{Taylor(1974)}]{taylor1974sensitive}
Taylor, J. 1974, Astronomy and Astrophysics Supplement Series, 15, 367

\bibitem[{{van Belle} \& {von Braun}(2009)}]{vanbelle2009directly}
{van Belle}, G.~T., \& {von Braun}, K. 2009, \apj, 694, 1085

\bibitem[{{van Leeuwen}(2007)}]{vanleeuwen2007hip}
{van Leeuwen}, F. 2007, \aap, 474, 653

\bibitem[{Wells {et~al.}(2018)Wells, Poppenhaeger, Watson, \&
  Heller}]{Wells2018}
Wells, R., Poppenhaeger, K., Watson, C.~A., \& Heller, R. 2018, Monthly Notices
  of the Royal Astronomical Society, 473, 345

\bibitem[{Wolszczan \& Frail(1992)}]{wolszczan1992planetary}
Wolszczan, A., \& Frail, D.~A. 1992, Nature, 355, 145

\bibitem[{{Worden} {et~al.}(2017){Worden}, {Drew}, {Siemion}, {Werthimer},
  {DeBoer}, {Croft}, {MacMahon}, {Lebofsky}, {Isaacson}, {Hickish}, {Price},
  {Gajjar}, \& {Wright}}]{Worden2017}
{Worden}, S.~P., {Drew}, J., {Siemion}, A., {et~al.} 2017, Acta Astronautica,
  139, 98

\bibitem[{Wright {et~al.}(2018{\natexlab{a}})Wright, Kanodia, \&
  Lubar}]{wright2018much}
Wright, J.~T., Kanodia, S., \& Lubar, E. 2018{\natexlab{a}}, The Astronomical
  Journal, 156, 260

\bibitem[{Wright {et~al.}(2018{\natexlab{b}})Wright, Sheikh, Alm{\'a}r,
  Denning, Dick, \& Tarter}]{wright2018recommendations}
Wright, J.~T., Sheikh, S., Alm{\'a}r, I., {et~al.} 2018{\natexlab{b}}, arXiv
  preprint arXiv:1809.06857

\bibitem[{{Zacharias} {et~al.}(2012){Zacharias}, {Finch}, {Girard}, {Henden},
  {Bartlett}, {Monet}, \& {Zacharias}}]{zacharias2012usno}
{Zacharias}, N., {Finch}, C.~T., {Girard}, T.~M., {et~al.} 2012, VizieR Online
  Data Catalog, I/322A

\end{thebibliography}

\end{document}